\documentclass[12pt]{article}
\usepackage{mathrsfs}

\usepackage{color}
\usepackage{multicol}
\usepackage{graphicx,natbib,amssymb}%,lineno}
\usepackage{amsmath, amsthm}
\usepackage{fancyhdr}
\usepackage{indentfirst}
\usepackage{geometry}
\usepackage{lineno}
\usepackage{epstopdf}

\numberwithin{equation}{section}
\theoremstyle{plain}
\usepackage[bf, small]{titlesec}
\titleformat{\section}{\bf\large}{\thesection.\,}{0.24em}{}
\titlespacing{\section}{0cm}{*1.5}{*1.1}
\titleformat{\subsection}{\it}{\thesubsection.\enspace}{0.5em}{}
\titlespacing{\subsection}{0.5cm}{*4}{*1.5}
\newtheoremstyle{theorem}%name
{10pt} % space above
{10pt} % space below
{\sl} % bofy font
{\parindent} % ident - empty=no indent, \parindent= paragraph indent
{\bf} % thm head font
{. } % punctuation after thm head
{ } % space after thm head: `` ``=normal \newline=linebreak
{} % thm head specification
\theoremstyle{theorem}

\newtheoremstyle{defi}%name
{10pt} % space above
{10pt} % space below
{\rm} % bofy font
{\parindent} % ident - empty=no indent, \parindent= paragraph indent
{\bf} % thm head font
{. } % punctuation after thm head
{ } % space after thm head: `` ``=normal \newline=linebreak
{} % thm head specification
\theoremstyle{defi}

\begin{document}

%\linenumbers

\title{Pareto-efficient biological pest control enable high efficacy at small costs}

\author{{Niklas L.P. Lundstr\"{o}m$^1$, Hong Zhang$^{1,2}$, {\AA}ke Br\"{a}nnstr\"{o}m$^{1,3}$}\\\\
\emph{\small $^1$ Department of Mathematics and Mathematical Statistics,} \\
\emph{\small Ume{\aa} University, Ume{\aa}, SE-901 87, Sweden}\\
\emph{\small niklas.lundstrom@umu.se (corresponding author)}\\\\
\emph{\small $^2$ Department of Mathematics,}\\
\emph{\small Jiangsu University, Zhenjiang, Jiangsu 212013, P.R. China}\\\\
%cnczzhanghong@163.com
%
\emph{\small $^3$ Evolution and Ecology Program,} \\
\emph{\small International Institute for Applied Systems Analysis,} \\
\emph{\small 2361 Laxenburg, Austria}
%ake.brannstrom@math.umu.se
}
\date{}
\maketitle
%\newpage
\begin{abstract}
Biological pest control is increasingly used in agriculture as a an alternative to traditional chemical pest control. In many cases, this involves a one-off or periodic release of naturally occurring and/or genetically modified enemies such as predators, parasitoids, or pathogens. As the interaction between these enemies and the pest is complex and the production of natural enemies potentially expensive, it is not surprising that both the efficacy and economic viability of biological pest control are debated. Here, we investigate the performance of very simple control strategies. In particular, we show how Pareto-efficient one-off or periodic release strategies, that optimally trade off between efficacy and economic viability, can be devised and used to enable high efficacy at small economic costs. We demonstrate our method on a pest-pathogen-crop model with a tunable immigration rate of pests. By analyzing this model, we demonstrate that simple Pareto-efficient one-off and periodic release strategies are efficacious and simultaneously have profits that are close to the theoretical maximum obtained by strategies optimizing only the profit. When the immigration rate of pests is low to intermediate, one-off control strategies are sufficient, and when the immigration of pests is high, periodic release strategies are preferable. The methods presented here can be extended to more complex scenarios and be used to identify promising biological pest control strategies in many circumstances.\\
\end{abstract}


%\newpage

%%%%%%%%%%%%%%%%%%%%%%%%%%%%%%%%%%%%%%%%%%%%%%%%%%%%%%%%%%%%%%%%%%%%%%
%%%%%%%%%%%%%%%%%%%%%%%%%%%%%%%%%%%%%%%%%%%%%%%%%%%%%%%%%%%%%%%%%%%%%%
%%%%%%%%%%%%%%%%%%%%%%%%%%%%%%%%%%%%%%%%%%%%%%%%%%%%%%%%%%%%%%%%%%%%%%
%%%%%%%%%%%%%%%%%%%%%%%%%%%%%%%%%%%%%%%%%%%%%%%%%%%%%%%%%%%%%%%%%%%%%%
%%%%%%%%%%%%%%%%%%%%%%%%%%%%%%%%%%%%%%%%%%%%%%%%%%%%%%%%%%%%%%%%%%%%%%
%%%%%%%%%%%%%%%%%%%%%%%%%%%%%%%%%%%%%%%%%%%%%%%%%%%%%%%%%%%%%%%%%%%%%%
%%%%%%%%%%%%%%%%%%%%%%%%%%%%%%%%%%%%%%%%%%%%%%%%%%%%%%%%%%%%%%%%%%%%%%
%%%%%%%%%%%%%%%%%%%%%%%%%%%%%%%%%%%%%%%%%%%%%%%%%%%%%%%%%%%%%%%%%%%%%%
%%%%%%%%%%%%%%%%%%%%%%%%%%%%%%%%%%%%%%%%%%%%%%%%%%%%%%%%%%%%%%%%%%%%%%
%%%%%%%%%%%%%%%%%%%%%%%%%%%%%%%%%%%%%%%%%%%%%%%%%%%%%%%%%%%%%%%%%%%%%%

\section{Introduction}

Pests are major concerns in agriculture. 
Local outbreaks cause financial losses and regional outbreaks threaten the food security of entire populations. 
This is of particular concern in developing
nations where agriculture constitutes a larger share of the economy
but in which agricultural practices have not yet reached the same technical and
procedural standards as in developed nations. In India, for example,
the ``Army worm'' \emph{Spodoptera litura} (Fabr.) has defoliated many
economically important crops including cotton, sunflower, and soybean
\citep{ref1}.  Farmers have traditionally resorted to pesticides to
prevent and mitigate pest outbreaks, but their use may have unwanted
consequences including insect resistance, resurgence, outbreak of
secondary pests, and pesticide residues affecting human health and the
environment. Indeed, heavy usage of synthetic pesticides have been
linked to pest resistance, pest resurgence, health risks from
exposure, and food contamination \citep{ref2+, ref2}

Biological pest control is an alternative to chemical pest control
%\citep{fle,lou}
in which naturally occurring enemies such as predators, parasitoids, or pathogens rather than
pesticides are used to control the pests.
The use of naturally occurring enemies to suppress insect pests has several advantages over chemical pest
control, in particular safety for farmers, consumers, and non-targeted
organisms. Biological pest control can
potentially be efficacious at low cost and should not normally pose any
danger for either farmers or consumers. They can be host-specific,
they preserve natural enemies, and they
may beneficially impact biodiversity \citep{lac}.
%(Bailey, 1975; Flint \& van den Bosch, 1981; Cory et al., 1994; Fuxa,
%2004).
Unlike the use of pesticides, there is little consensus on how to apply
biological control for maximal efficiency.
%This is not only because biological control agents have been introduced relatively recently,
%but it is also a consequence of a
One reason for this is the complex interplay of non-linear
interactions between the crop, the pest, and the natural enemy. The
potential benefits of improved biological pest-control strategies are
particularly large for inundative and augmentative applications, 
in which large numbers of natural enemies are released, as the timing of the
release may significantly affect the total cost and efficacy.

%% microbial control to succeed, the dynamics of pests and pest pathogens
%% or viruses should be understood \citep{and2}, which help researchers
%% approximately carry out the computational experiments of the real
%% system, operating only a model of the system with major parameters,
%% which should be observed. In addition, they can make use of
%% mathematical results to aid the consummation of field experiments
%% (\cite{bri1,bri2,con,preedy}).

% Tang och Chang har studerat impulsive pest control, see Cardoso et al. (2009)

To the authors knowledge, 
a handful of studies have explored design of biological pest-control
strategies from the perspective of mathematical analysis and/or
optimal control theory. These studies have considered problems of
bioeconomic equilibrium, demographic stability, and optimal-release
strategies \citep{get,grasman,bha,raf,card}. While these studies have
furthered our understanding of biological pest control, the proposed
pest-control strategies may not easily be communicated to agriculture
professionals as they typically lack a regular pattern and sometimes
require continuous release of natural enemies. Moreover, with
%\citet{card}
Cardoso et al. (2009) as an important exception, only single-objective
optimization is usually considered. Finally, to the authors knowledge, the studies to
date have not explicitly modeled the crop, which as a third dynamic
state variable could potentially impact the results. Developing simple
but efficient rules for biological pest control in agricultural
systems with crop-pest-enemy interactions thus remains an important
challenge from both a theoretical and applied perspective.

In this paper, we suggest a simple method for developing strategies for biological pest control that
are easy to apply, efficacious, and simultaneously near optimal in terms
of profit. The strategies are Pareto-efficient in that they optimally trade off between profit and efficacy. 
We demonstrate our method on a dynamic model of the
\emph{Spodoptera litura} worm defoliating soybean crops while being
controlled by a natural enemy, the \emph{Spodoptera polyhedrosis}
virus \citep{cherry, fuxa}.  Specifically, we investigate one-off
control strategies and periodic control strategies.  Using our measures
of efficacy and profit, we find Pareto-efficient one-off and
periodic control strategies that are close to optimal in the sense
of profit and simultaneously not sensitive to perturbations.  We show
that one-off control strategies are preferable when immigration of
pests is relatively low to intermediate. We also show that, for high
immigration rates, one-off control can be replaced by simple periodic
controls to achieve even better results.

%%%%%%%%%%%%%%%%%%%%%%%%%%%%%%%%%%%%%%%%%%%%%%%%%%%%%%%%%%%%%%%%%%%%%%
%%%%%%%%%%%%%%%%%%%%%%%%%%%%%%%%%%%%%%%%%%%%%%%%%%%%%%%%%%%%%%%%%%%%%%
%%%%%%%%%%%%%%%%%%%%%%%%%%%%%%%%%%%%%%%%%%%%%%%%%%%%%%%%%%%%%%%%%%%%%%
%%%%%%%%%%%%%%%%%%%%%%%%%%%%%%%%%%%%%%%%%%%%%%%%%%%%%%%%%%%%%%%%%%%%%%
%%%%%%%%%%%%%%%%%%%%%%%%%%%%%%%%%%%%%%%%%%%%%%%%%%%%%%%%%%%%%%%%%%%%%%
%%%%%%%%%%%%%%%%%%%%%%%%%%%%%%%%%%%%%%%%%%%%%%%%%%%%%%%%%%%%%%%%%%%%%%
%%%%%%%%%%%%%%%%%%%%%%%%%%%%%%%%%%%%%%%%%%%%%%%%%%%%%%%%%%%%%%%%%%%%%%
%%%%%%%%%%%%%%%%%%%%%%%%%%%%%%%%%%%%%%%%%%%%%%%%%%%%%%%%%%%%%%%%%%%%%%
%%%%%%%%%%%%%%%%%%%%%%%%%%%%%%%%%%%%%%%%%%%%%%%%%%%%%%%%%%%%%%%%%%%%%%
%%%%%%%%%%%%%%%%%%%%%%%%%%%%%%%%%%%%%%%%%%%%%%%%%%%%%%%%%%%%%%%%%%%%%%

\section{Model}
\label{sec:model}

In this section,
we first present the sample model on which we will demonstrate our method for deriving simple control strategies.
This model consist of a pest-pathogen-crop system in which the pest is controlled biologically
through the release of individuals that are infected with a virus.
The infection spreads into the susceptible pest population and thus control the growth of pest biomass in the field.
Second, we give basic results on the dynamics of the model considering equilibria and their stability in terms of their basin of attraction.

%%%%%%%%%%%%%%%%%%%%%%%%%%%%%%%%%%%%%%%%%%%%%%%%%%%%%%%%%%%%%%%%%%%%%%
%%%%%%%%%%%%%%%%%%%%%%%%%%%%%%%%%%%%%%%%%%%%%%%%%%%%%%%%%%%%%%%%%%%%%%
%%%%%%%%%%%%%%%%%%%%%%%%%%%%%%%%%%%%%%%%%%%%%%%%%%%%%%%%%%%%%%%%%%%%%%

\subsection{Crop-pest-pathogen system}
\label{sec:modelling_crop}

We model a crop-pest-pathogen system inspired by soybeans devoured by the
``army worm" \emph{Spodoptera litura} \citep{kun}. The army worm is
being controlled biologically through the release of individuals that
are infected with \emph{Spodoptera polyhedrosis} virus \citep{orei}.
The biomass of soybean is denoted $C = C(t)$, while the density of
infected and susceptible pests are respectively denoted $P_I = P_I(t)$ and
$P_S = P_S(t)$.  Disease transmission between susceptible and infected
pests follow the law of mass action with a constant transmission
coefficient $\beta$ \citep{mcc}.
An overview of model variables and parameters is given in Table~\ref{tab01}.

To arrive at a tractable model that incorporates the essential features
of the crop-pest-pathogen system, we integrate two influential and established
models in theoretical biology, the Rosenzweig-MacArthur predator-prey model \citep{RMA} and the
classical SI-compartment model in epidemiology \citep{SI}. On this basis, we assume that
the dynamics of the crop are given by the following ordinary differential equation:
\begin{align}\label{eq:syst_eq_1}
\frac{{\rm d}\,C}{{\rm d}\,t} \,=\, r\, C \left(1\,-\,\frac{C}{K}\right)\,-\,\frac{a_S\,C\,P_S}{b_S\,+\,C}\,-\,\frac{a_I\, C\, P_I}{b_I\,+\,C},
\end{align}
where the terms on the right hand side represent logistic growth in the
absence of consumption by the pest, consumption by susceptible pests, and
consumption by infected pests, respectively.
The dynamics of susceptible and infected pests are respectively given
by:
\begin{align}\label{eq:syst_eq_2}
  \frac{{\rm d}\,P_S}{{\rm d}\,t}\, =\, c_S\, \frac{a_S\, C\, P_S}{b_S\,+\,C} \,  -\,   \beta\, P_S\,P_I \,  -\,   d_S\, P_S\,   + \,  A,
\end{align}
\begin{align}\label{eq:syst_eq_3}
  \frac{{\rm d}\,P_I}{{\rm d}\,t}\, = \,c_I\, \frac{a_I\, C \,P_I}{b_I\,+\,C} \,  +  \, \beta \,P_S\,P_I \,  - \,  d_I \,P_I.
\end{align}
From left to right, the terms represent reproduction of pests, disease transmission and mortality.
Susceptible pests are assumed to immigrate from neighboring fields at a constant rate $A \geq 0$.
Both susceptible and infected pests are capable of crop consumption and reproduction. 
Virulence is assumed to affect infected pests through reduced fecundity, 
reduced crop consumption rate, and increased mortality. 
These assumptions are reflected in the following
conditions on the parameters,
$a_S \gg a_I$,
$c_S \gg c_I$ and
$d_S < d_I$.
%$b_S > b_I$.
In Table~\ref{tab01} we state units and numerical values for all model parameters 
%
%While the techniques developed in this paper are independent of this special parametrization,
%the values of the biological parameters
%are
based on published papers
\citep{ball, R-N, xiaov, dale_1, nyref1, nyref2}.
%Ball et al. (2000),
%Dale (2006),
%Ruis-Nogueira et al. (2001),
%Liao et al (2016),
%Liu et al. (2015),
%Xiao and Van Den Bosch (2003),
While the techniques developed in this paper are independent of this special parametrization,
the determined control strategies do depend on our chosen parametrization. 
To strengthen our results beyond our particular parameter values we perform a robustness check in Section~\ref{sect_robus}.

Before discussing basic dynamical properties of system
\eqref{eq:syst_eq_1}--\eqref{eq:syst_eq_3},
we note that structurally similar mathematical systems have been used by 
%\citep{LXW10}
Li et al. (2010) to study a predator-prey system with group defense and impulsive control strategies,
and by Zhang and Georgescu (2015)
%\citep{ZG15}
to study the influence of the multiplicity of infection upon the dynamics
of a crop-pest-pathogen model with defense mechanisms.
%
% TABLE 1
%

\begin{table}[!hbp]
\caption{State variables and parameters of the crop-pest-pathogen model.
%In the table ``-'' means that the impact of the parameter value will be explored in Sections~\ref{sec:model_dynamics} and \ref{sec:results}.
%\emph{Sources}:
%(1) %\cite{ball};
%Ball et al. (2000);
%(2)
%Ruis-Nogueira etal. (2001);
%(3) %\cite{xiaov};
%Xiao and Van Den Bosch (2003);
%(4) %\cite{dale_1};
%Dale (2006);
%(5)
%U.S. Department of Agriculture's National Agricultural Statistics Service.
}
\label{tab01}
\begin{center}
\footnotesize{\begin{tabular}{@{}llcll}
\hline
Quantity  & Symbol & Value  & Unit 
\\ \hline
Biomass of the soybean crop population          & $C$                     & variable             & gram\ m$^{-2}$                 \\ \hline
Density of the susceptible pest population      & $P_S$                   & variable             & m$^{-2}$                 \\ \hline
Density of the infected pest population         & $P_I$                   & variable             & m$^{-2}$                 \\ \hline
Crop intrinsic growth rate                      & $r$                     & 0.45                 & day$^{-1}$                     \\ \hline
Carrying capacity of the soybean crop           & $K$                     & 500                  & gram\ m$^{-2}$                 \\ \hline
Consumption rate of susceptible pests           & $a_S$                   & 0.8                  & gram day$^{-1}$
\\ \hline
Consumption rate of infected pests              & $a_I$                   & 0.01                 & gram day$^{-1}$
\\ \hline
Half saturation constant of susceptible pests   & $b_S$                   & 200                  & gram m$^{-2}$                  \\ \hline
Half saturation constant of infected pests      & $b_I$                   & 50                   & gram m$^{-2}$                  \\ \hline
Reproductive rate of susceptible pests          & $c_S$                   & 0.5                  & gram$^{-1}$               \\ \hline
Reproductive rate of infected pests             & $c_I$                   & 0.01                 & gram$^{-1}$               \\ \hline
Mortality rate of susceptible pests             & $d_S$                   & 0.1                  & day$^{-1}$                    \\ \hline
Mortality rate of infected pests               & $d_I$                   & 0.8                  & day$^{-1}$                     \\ \hline
Contact rate                                    & $\beta$                 & 0.008                & m$^{2}$ day$^{-1}$
\\ \hline
Immigration rate of susceptible pests           & $A$                     & -                    & m$^{-2}$ day$^{-1}$       \\ \hline
Length of the growth season             		   & $t_{\text{season}}$      & 140                  & day                           \\ \hline
Initial biomass of soybeans                     & $C(0)$                   & 5                    & gram m$^{-2}$                  \\ \hline
Initial amount of susceptible pests             & $P_{S}(0)$              & 0                    & m$^{-2}$                  \\ \hline
Initial amount of infected pests           	   & $P_{I}(0)$              & -                    & m$^{-2}$                  \\ \hline
Price of soybeans                               & $p_{\text{crop}}$       & $4.5\times 10^{-4}$  & \$\ gram$^{-1}$               \\ \hline
Fixed other costs                               & $p_{\text{fixed}}$      & $0.01$               & \$\ m$^{-2}$                  \\ \hline
Price per infected pests                        & $p_{\text{infected}}$   & $2\times 10^{-5}$    & \$
\\ \hline
Price of placing infected pests in the field      & $p_{\text{labour}}$       & $5\times 10^{-3}$    & \$\ m$^{-2}$                   \\ \hline
\end{tabular}}
\end{center}
\end{table}

\subsection{Model dynamics}
\label{sec:model_dynamics}

\begin{figure}
\begin{center}
  \includegraphics[scale=0.40]{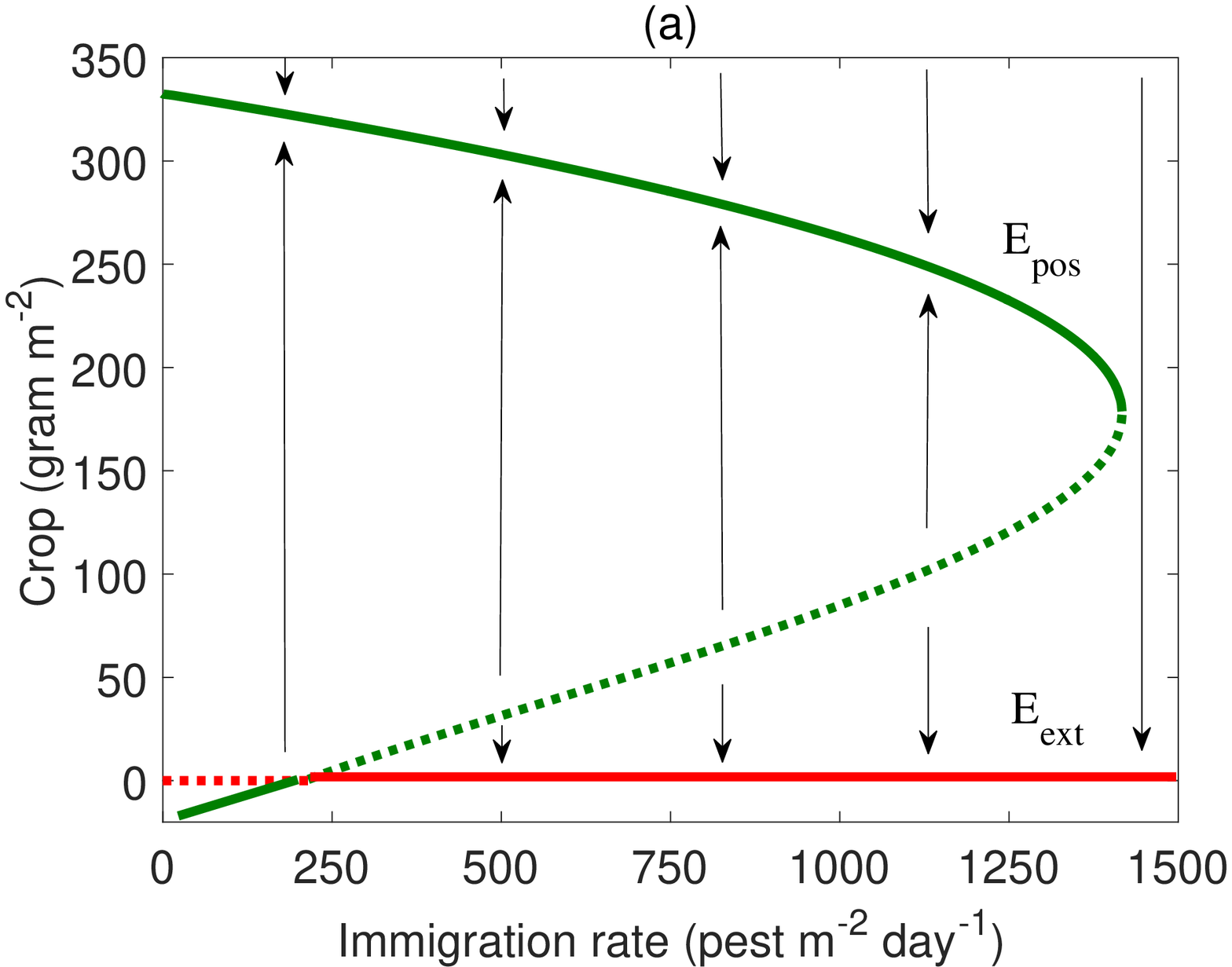}\hspace{0.0cm}
  \includegraphics[scale=0.40]{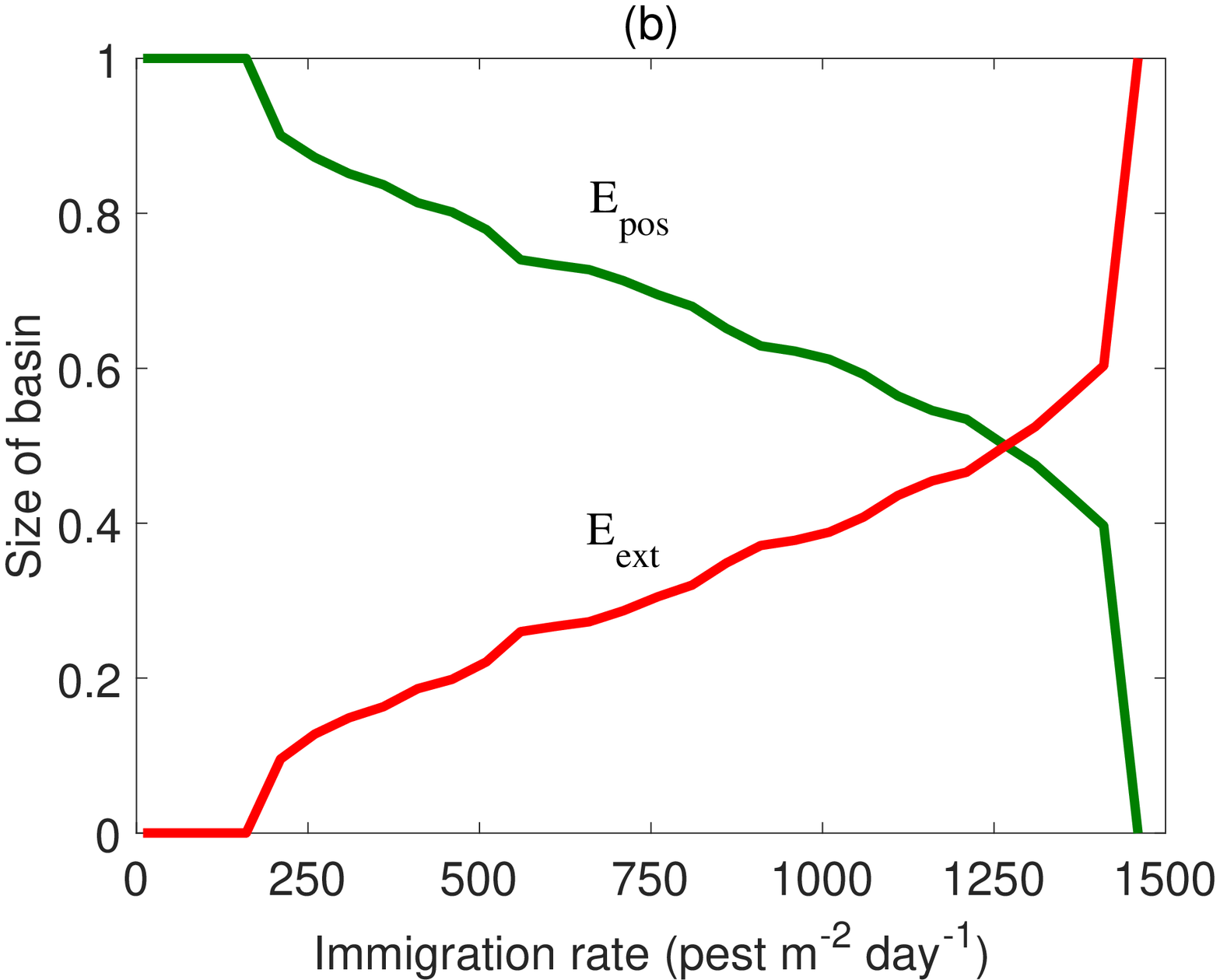}
\caption{(a) Bifurcation diagram showing the equilibria $E_{\text{pos}}$ and $E_{\text{ext}}$
as a function of immigration of susceptible pests $A$.
Solid curves denote stable equilibria while dotted curves denote unstable equilibria.
(b) Size of basin of the equilibria $E_{\text{pos}}$ and $E_{\text{ext}}$
as a function of immigration of susceptible pests $A$.
The extinction equilibria $E_{\text{ext}}$ attracts more and more of the state space as $A$ increases.}
\label{fig:basin}
\end{center}
\end{figure}

System \eqref{eq:syst_eq_1}-\eqref{eq:syst_eq_3} always has an extinction equilibrium
$(E_{\text{ext}}$, a crop free state) given by
\begin{align}\label{eq:ext}
C \,=\, 0, \quad P_S \,=\, \frac{d_I}{\beta} \quad \text{and} \quad P_I \,=\, \frac{A}{d_I} \,-\, \frac{d_S}{\beta}.
\end{align}
The eigenvalues of the Jacobian matrix of system \eqref{eq:syst_eq_1}-\eqref{eq:syst_eq_3} at $E_{\text{ext}}$ are
\begin{align*}
\lambda_{1,2}\, =\,  -\,  \frac{A\, \beta\, \pm \,\sqrt{A^2\, \beta^2 \,-\, 4\, A\, \beta\, d_I^2\, +\, 4\, d_I^3\, d_S}}{2\, d_I}, \quad
\lambda_3\, =\, r\, - \,\frac{A \, a_I}{d_I\, b_I} \,+\, \frac{a_I\, d_S}{\beta\, b_I}\, -\,\frac{a_S\, d_I}{\beta\, b_S}.
\end{align*}
From this we conclude that the extinction equilibrium $E_{\text{ext}}$ becomes stable
in a transcritical bifurcation at $A = 210$ as the
third eigenvalue passes through the origin and becomes negative for $A > 210$.
When the immigration rate is relatively low, $0 < A < 210$,
simulations using MATLABs ODE-solver ode45
indicate that system \eqref{eq:syst_eq_1}--\eqref{eq:syst_eq_3} has
only one attractor which is a positive globally stable (with respect to the positive state space) equilibrium ($E_{\text{pos}}$).
For higher immigration rates,  $210 < A < 1417$,
$E_{\text{ext}}$ is also stable and the system therefore has two attractors simultaneously.
At $A = 1417$ the positive stable equilibrium $E_{\text{pos}}$ collides with its unstable branch and disappears in a fold bifurcation,
and for $A>1417$ then $E_{\text{ext}}$ is the global attractor,
see Fig.~\ref{fig:basin}a.
%If the initial biomass of soybeans is small while immigration is high,
%then the extinction equilibrium is the attractor which lead to the crop-free state.

If $E_{\text{ext}}$ has a large basin of attraction (henceforth basin) while $E_{\text{pos}}$
has a small basin,
a perturbation may cause the solution to jump from $E_{\text{pos}}$ to $E_{\text{ext}}$ with a high probability, resulting in crop extinction.
Hence, the size of basin is important and
to understand stability beyond a local analysis we proceed along the lines of Lundstr\"om and Aidanp\"a\"a (2007)
by considering the size of basin as a stability measure.
In particular, we start a trajectory at each of the 
$11^3$ initial conditions $\left(C(0), P_S(0), P_I(0)\right)$ given by points in the set 
$\{1,50,100,\dots,500\} \times \{1,30,60,\dots, 300\}\times \{1,30,60, \dots, 300\}$ 
and integrate the system until trajectories reach a small neighborhood of either 
$E_{\text{pos}}$ or $E_{\text{ext}}$.
Letting $N_{\text{pos}}$ and $N_{\text{ext}}$ denote the number of initial conditions that reached $E_{\text{pos}}$  and $E_{\text{ext}}$, respectively,
we get an estimate of the size of the basins by
\begin{align*}
\text{size of basin}\,(E_{\text{pos}})\, \approx\, \frac{N_{\text{pos}}}{N_{\text{tot}}}, \quad
\text{size of basin}\,(E_{\text{ext}})\, \approx\, \frac{N_{\text{ext}}}{N_{\text{tot}}},
\end{align*}
where $N_{\text{tot}}$ denotes the total number of tested initial conditions.

Figure~\ref{fig:basin}b shows the size of basin of both equilibria as a function of immigration of susceptible pests, $A$.
As $A$ increases, the stability in sense of basin of the positive equilibrium $E_{\text{pos}}$ decreases as more and more of the positive state space is attracted by the extinction equilibrium $E_{\text{ext}}$.
It will therefore be more and more important to keep the crop biomass away from $E_{\text{ext}}$ as immigration of susceptible pests increases.
Since the curves in Fig.~\ref{fig:basin}b sum to one,
all tested initial conditions end up either at equilibrium $E_{\text{pos}}$ or $E_{\text{ext}}$,
giving strong arguments for the fact that these equilibria are the only existing attractors in 
system \eqref{eq:syst_eq_1}--\eqref{eq:syst_eq_3}.

%
% FIGURE 0
%

\begin{figure}
\begin{center}
  \includegraphics[scale=0.40]{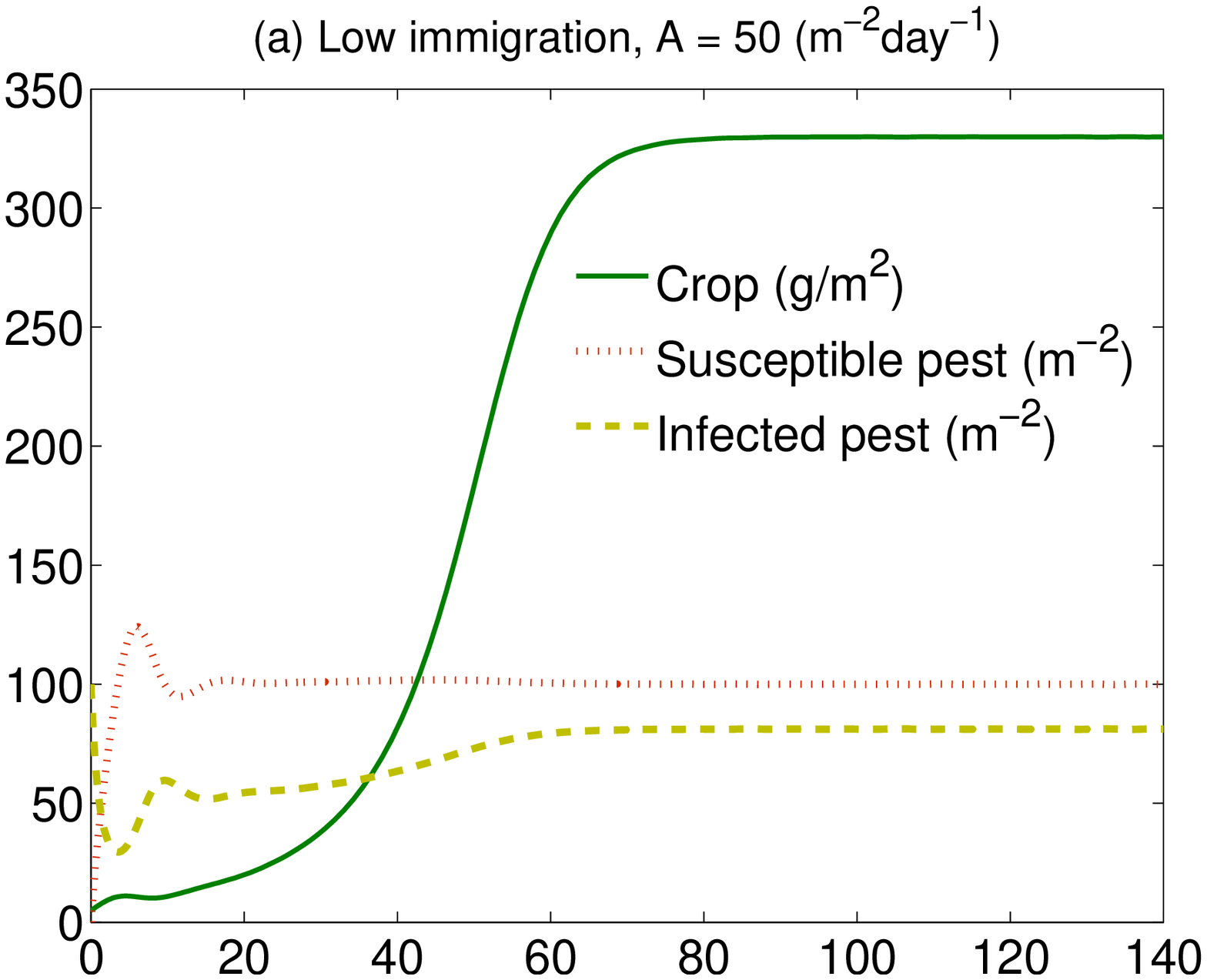}\hspace{0.6cm}
  \includegraphics[scale=0.40]{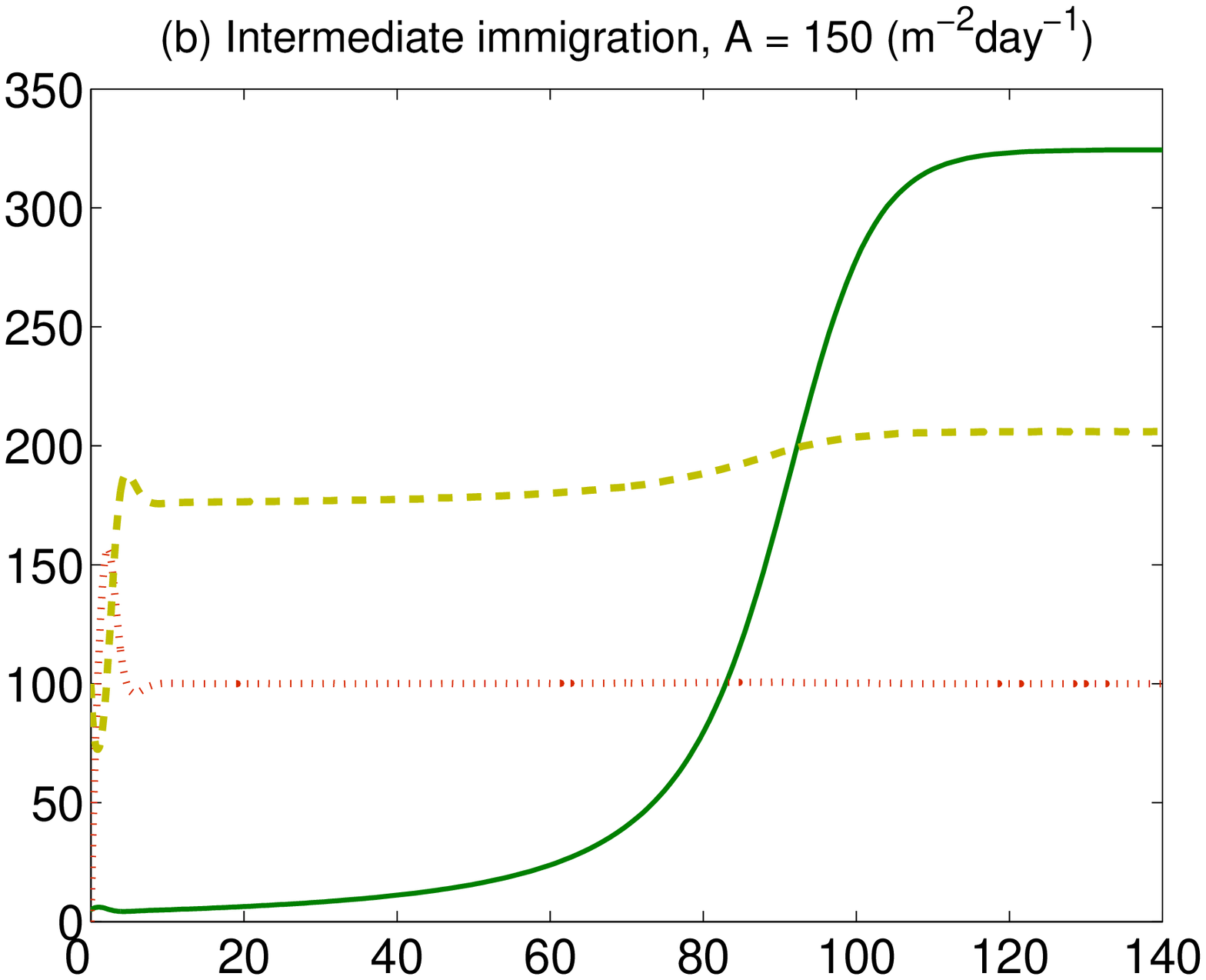}
  \includegraphics[scale=0.40]{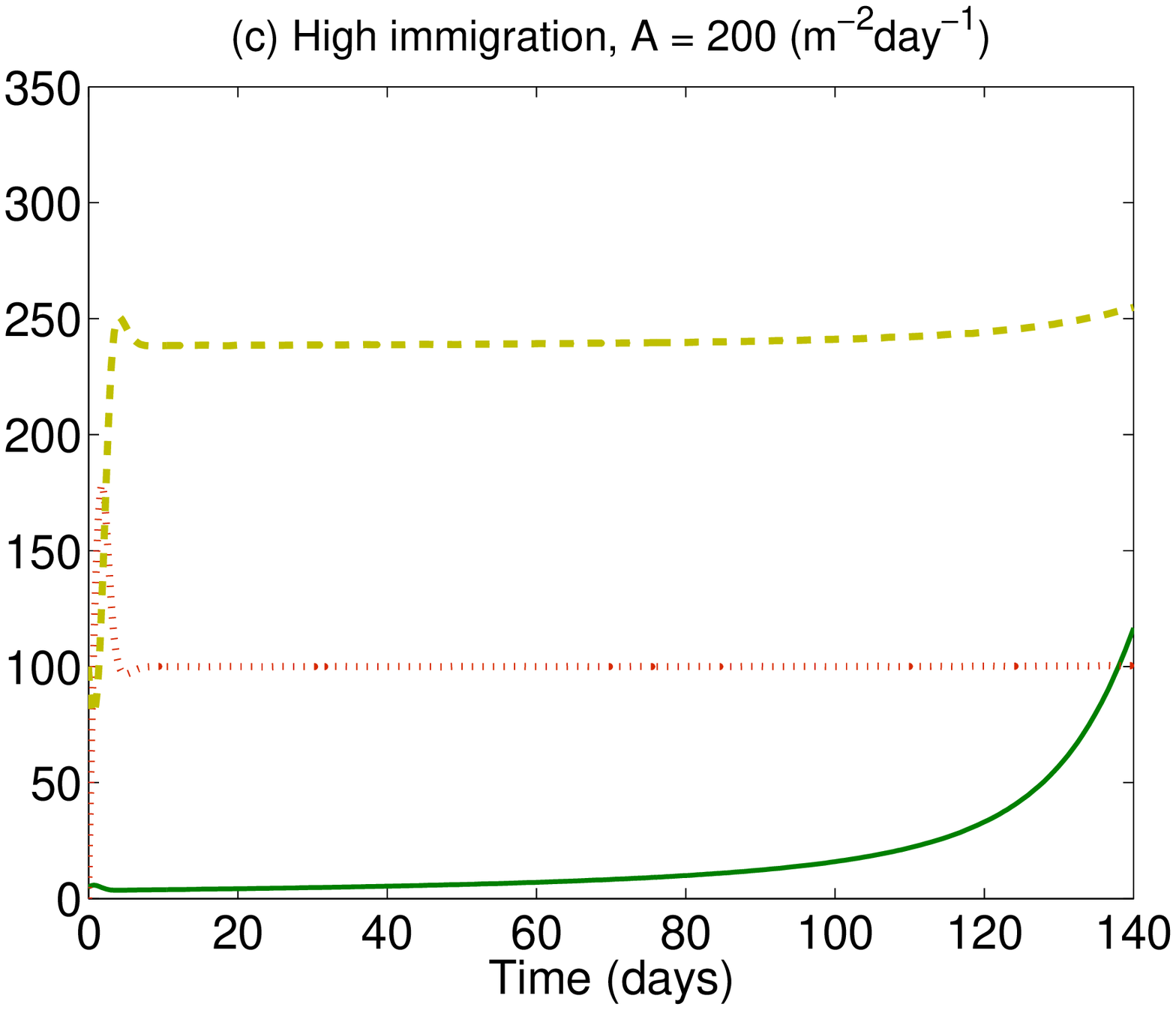}\hspace{0.3cm}
  \includegraphics[scale=0.40]{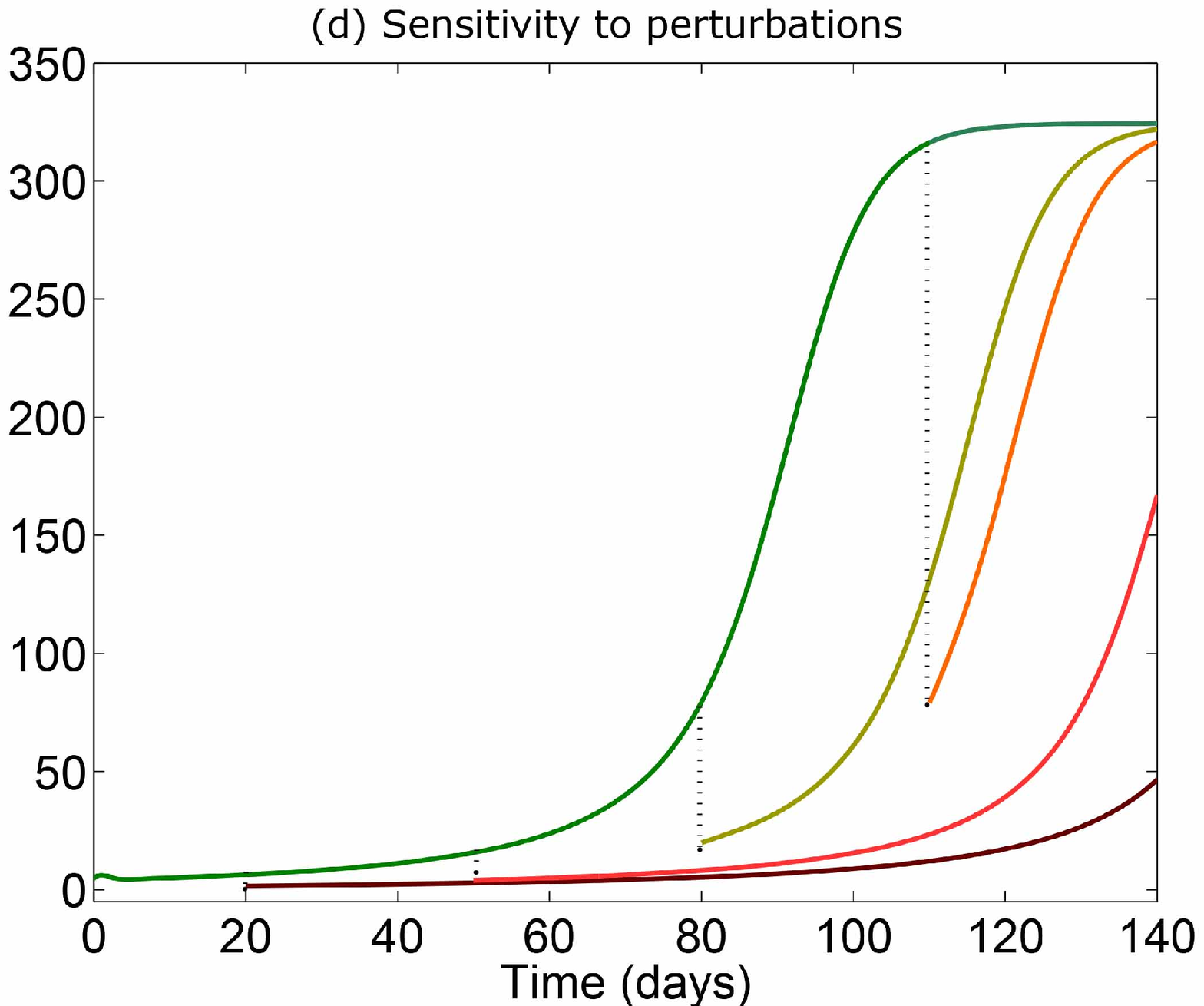}
\caption{(a)-(c): Trajectories of crop (green, solid), susceptible pests (red, dotted) and infected pests (yellow, dashed).
The soybean crop biomass increases rapidly at near half its equilibrium value.
With higher immigration rate of susceptible pests, 
the rapid increase comes later in the season.
In panel (c) it is delayed beyond the end of the season.
(d): The final crop biomass is sensitive to perturbations
in the beginning of the season when the crop biomass is small and grows slowly.
Trajectory of crop biomass without perturbation (green).
Trajectories of crop biomass with a sudden reduction to 1/4 of its amount at time
110 (yellow),
80 (orange),
50 (red)
and 20 (dark red).
Initial amount of infected pests is set to $P_I(0) = 100$ g/m$^2$ 
and the immigration rate of susceptible pests is $A = 150$.}
\label{fig:one}
\end{center}
\end{figure}

Typically, the system behave in a way that is shown by the trajectories in Fig.~\ref{fig:one}.
The crop population will start growing slowly,
followed by a rapid increase due to the logistic growth assumption.
As the crop biomass is small it grows slowly, and a small perturbation
may cause it to stay small during a long time interval,
perhaps too long in order to get a good yield at the end of the growth season.
If immigration is relatively high, a perturbation may also result in that
the crop dies due to the attracting extinction equilibria $E_{\text{ext}}$.
These facts motivate our efficacy measure \textit{half-biomass time} introduced in Section~\ref{sec:dual}.
From Fig.~\ref{fig:one}a-c we can also see that as immigration of susceptible pests increases,
then the disease spreads in the pest population so that the infected pests biomass $P_I$,
rather than susceptible pests $P_S$, increases.
When crop biomass is small, this can be understood by perturbing off the expression
of the extinction equilibrium in \eqref{eq:ext}. In particular,
observe the term $A/d_I$ in the expression for $P_I$.

% FIGURE 1
%

As the system \eqref{eq:syst_eq_1}--\eqref{eq:syst_eq_3} is based on a predator-prey model of Rosenzweig-MacArthur type,
one may expect it to have limit cycles in some cases.
However, the extensive simulations of the basin of attraction presented
in Fig.~\ref{fig:basin}b indicate no attracting cycles.
In Section~\ref{sect_robus} we consider robustness of our results with respect to variations in the parameter values, and these simulations also agree that the above equilibria remain the only attractors.

%%%%%%%%%%%%%%%%%%%%%%%%%%%%%%%%%%%%%%%%%%%%%%%%%%%%%%%%%%%%%%%%%%%%%%
%%%%%%%%%%%%%%%%%%%%%%%%%%%%%%%%%%%%%%%%%%%%%%%%%%%%%%%%%%%%%%%%%%%%%%
%%%%%%%%%%%%%%%%%%%%%%%%%%%%%%%%%%%%%%%%%%%%%%%%%%%%%%%%%%%%%%%%%%%%%%

\section{Control strategies}
\label{sec:model_control}

Next, we introduce and give a precise definition of the control strategies that we consider for the release of
infected individuals.
We chose our class of control strategies for conceptual simplicity,
though as we will show these strategies are capable of achieving near-optimal profits.
Before describing the control strategies, we briefly note that the sample model in
\eqref{eq:syst_eq_1}-\eqref{eq:syst_eq_3} considers only the number of pests as well as biomass of crop,
no spatial dependence is involved.
Consequently, we do not discuss practical methods for release of the infected pests in order to
efficiently and inexpesively distribute them evenly across the field.

Among the simple control strategies, 
we first consider releasing infected pests only once,
at the beginning of the season.
We call this strategy \textit{one-off control}.
This simple strategy has one control variable,
the total amount of released infected pests $\widetilde P_I$,
and is specified by the initial amount of infected pests as
$$
P_I(0) \,=\, \widetilde P_I.
$$

We next consider periodic control in which we assume that the same amount of infected pests
are released the first day of each week throughout the growing season.
Since the growth of the crop biomass is most sensitive to perturbations in the beginning of the season,
we always assume that the first release happens at the first day of the season.
The first release is thus implemented by setting the initial amount of infected pests, $P_I(0)$.
Then the same amount of infected pests is released the first day of any following week,
ending with week $N$, where $N = 1,2,3,\dots$.
Thus, $N = 1$ corresponds to \textit{one-off control}.
The periodic control strategy involves two control variables,
the number of weeks $N$ where an impulse of infected pests is released,
and the total amount of released infected pests $\widetilde{P}_I$.
In particular, the implementation of periodic control can formally be written,
for $N = 2,3,4,\dots$, as
$$
P_I(0) \,=\, \frac{\widetilde P_I}{N} \quad \text{and} \quad
P_I(s) \,\to\, P_I(s)\, + \, \frac{\widetilde P_I}{N},
$$
where $s=\{7,14,\dots 7(N-1)\}$.
%We also run simulations to find the optimal control function, considering profit as the only target under optimization.
%Here we use the software TOMLAB to find this control function giving this optimal profit.

%%%%%%%%%%%%%%%%%%%%%%%%%%%%%%%%%%%%%%%%%%%%%%%%%%%%%%%%%%%%%%%%%%%%%%
%%%%%%%%%%%%%%%%%%%%%%%%%%%%%%%%%%%%%%%%%%%%%%%%%%%%%%%%%%%%%%%%%%%%%%
%%%%%%%%%%%%%%%%%%%%%%%%%%%%%%%%%%%%%%%%%%%%%%%%%%%%%%%%%%%%%%%%%%%%%%

\section{Dual-objective approach}
\label{sec:dual}

We here define our measures of profit and efficacy,
after which we describe the concept of Pareto efficiency used
for dual-objective optimization.

Besides trying to optimize profit we also consider maximizing efficacy,
i.e. minimizing sensitivity to perturbations on the profit.
We will now define our profit function,
our measure of efficacy, i.e. half-biomass time,
and also recall the economic concept of Pareto efficiency,
which we will use to trade-off between the two objectives profit and efficacy.

\subsection{Profit measure}

To determine a profit function,
we first assume that yield is proportional to crop biomass and,
without loss of generality,
that the constant of proportionality is 1.
If $p_{\text{crop}}$ is the market price for the crop,
then the revenue is given by $p_{\text{crop}} \, C(t_{\text{season}})$ where $t_{\text{season}}$ is the time at the end of the growth season.
We also assume that the total cost is given by $p_{\text{fixed}}\, +\, p_{\text{infected}} \, \widetilde{P}_I\, +\, p_{\text{labour}} \, N$,
where $p_{\text{fixed}}$ represents annually-recurring costs for seed, sowing, fertilizer, irrigation, pest monitoring, taxes, etc. that remain fixed within a single growth season;
$p_{\text{infected}}$ is the price of infected pests;
and $p_{\text{labour}}$ is cost for work needed to place a burst of infected pests in the field.
Hence, the profit is given by
\begin{align}\label{eq:profit-func}
\textrm{Profit} \,=\, p_{\text{crop}} \, C(t_{\text{season}}) \,-\, p_{\text{fixed}} \,-\, p_{\text{infected}} \, \widetilde{P}_I \,-\, p_{\text{labour}}\, N.
\end{align}

%%%%%%%%%%%%%%%%%%%%%%%%%%%%%%%%%%%%%%%%%%%%%%%%%%%%%%%%%%%%%%%%%%%%%%
%%%%%%%%%%%%%%%%%%%%%%%%%%%%%%%%%%%%%%%%%%%%%%%%%%%%%%%%%%%%%%%%%%%%%%
%%%%%%%%%%%%%%%%%%%%%%%%%%%%%%%%%%%%%%%%%%%%%%%%%%%%%%%%%%%%%%%%%%%%%%

\subsection{Efficacy measure}

To construct a measure of efficacy on the final crop biomass we first consider the typical behavior of trajectories.
Figure~\ref{fig:one} shows that,
starting from a small initial crop biomass of 5 g/m$^2$,
then crop grows slowly in the beginning of the season followed by
a rapid increase starting at approximately 50 g/m$^2$.
In particular, when crop biomass is small,
then from \eqref{eq:syst_eq_1} we see that
$
%\frac{d C }{d t}
d C / d t \approx r C
$
and so the relative (to biomass) growth rate is constant.
Low profit results when the rapid increase in crop biomass takes place too late in the season.
Hence, it is important that the period of fast growing crop biomass is within the growth season,
as it is in Figs.~\ref{fig:one}a and b, but not in c.
As Fig.~\ref{fig:one}d shows,
the location of the rapid increase,
and hence the final crop biomass,
is sensitive to perturbations
in the beginning of the season when the crop biomass is small and grows slowly.
For relatively high immigration rates a perturbation on a small crop biomass may even cause the
crop to go extinct because of the attraction from the extinction equilibrium,
see Section~\ref{sec:model_dynamics}.
Hence, it is important to ensure that the crop comes away quickly from very small biomasses.
Based on these observations,
we chose to measure efficacy with \textit{half-biomass time},
which we define as the time needed for the crop
biomass to grow from its initial state to half of the equilibrium crop biomass.
This equilibrium may be thought of as an expected value of the final crop biomass.
The higher the half-biomass time is,
the larger is the risk for the farmer to obtain a small (or zero) final crop biomass
and hence a small (or negative) profit.

%%%%%%%%%%%%%%%%%%%%%%%%%%%%%%%%%%%%%%%%%%%%%%%%%%%%%%%%%%%%%%%%%%%%%%
%%%%%%%%%%%%%%%%%%%%%%%%%%%%%%%%%%%%%%%%%%%%%%%%%%%%%%%%%%%%%%%%%%%%%%
%%%%%%%%%%%%%%%%%%%%%%%%%%%%%%%%%%%%%%%%%%%%%%%%%%%%%%%%%%%%%%%%%%%%%%

\subsection{Pareto efficiency}

A control strategy is said to be Pareto efficient if any change in the control strategy
will make either the profit or the efficacy worse.
The Pareto front is defined as the set of all Pareto-efficient strategies.
Hence, the Pareto front consist of the ``best'' strategies and the choice of strategy on this front depends on
the desired trade-off between the two objectives. For further information on Pareto efficiency and dual-objective
optimization, we recommend an introductory textbook such as %\citep{K99}.
Karpagam (1999).
%%%%%%%%%%%%%%%%%%%%%%%%%%%%%%%%%%%%%%%%%%%%%%%%%%%%%%%%%%%%%%%%%%%%%%
%%%%%%%%%%%%%%%%%%%%%%%%%%%%%%%%%%%%%%%%%%%%%%%%%%%%%%%%%%%%%%%%%%%%%%
%%%%%%%%%%%%%%%%%%%%%%%%%%%%%%%%%%%%%%%%%%%%%%%%%%%%%%%%%%%%%%%%%%%%%%
%%%%%%%%%%%%%%%%%%%%%%%%%%%%%%%%%%%%%%%%%%%%%%%%%%%%%%%%%%%%%%%%%%%%%%
%%%%%%%%%%%%%%%%%%%%%%%%%%%%%%%%%%%%%%%%%%%%%%%%%%%%%%%%%%%%%%%%%%%%%%
%%%%%%%%%%%%%%%%%%%%%%%%%%%%%%%%%%%%%%%%%%%%%%%%%%%%%%%%%%%%%%%%%%%%%%
%%%%%%%%%%%%%%%%%%%%%%%%%%%%%%%%%%%%%%%%%%%%%%%%%%%%%%%%%%%%%%%%%%%%%%
%%%%%%%%%%%%%%%%%%%%%%%%%%%%%%%%%%%%%%%%%%%%%%%%%%%%%%%%%%%%%%%%%%%%%%
%%%%%%%%%%%%%%%%%%%%%%%%%%%%%%%%%%%%%%%%%%%%%%%%%%%%%%%%%%%%%%%%%%%%%%
%%%%%%%%%%%%%%%%%%%%%%%%%%%%%%%%%%%%%%%%%%%%%%%%%%%%%%%%%%%%%%%%%%%%%%

\section{Results}
\label{sec:results}

Having introduced our modeling framework, we now demonstrate how to find preferable control strategies.
First, we conclude that only a small reduction in profit allows for efficacious control strategies.
Second, we show that one-off control strategies are sufficient when immigration rates of susceptible pests are low to intermediate,
while periodic control strategies are recommended for high immigration rates.
Finally, we conclude that the determined control strategies are not far from optimal in terms of profit, despite their simplicity.
All simulations shown in Sections~\ref{sec:one-off}, \ref{sec:periodic} and \ref{sect_robus} are performed using MATLABs ODE-solver ode45, while in Section~\ref{sect_optimal} we used in addition the optimization software TOMLAB \citep{KK99}.

\subsection{A small reduction in profit allows efficacious control strategies}
\label{sec:one-off}

Figure~3a-c shows different one-off and periodic control strategies for low, 
intermediate and high immigration of susceptible pests.
In each panel, different curves correspond to different $N$, 
starting with one-off control, i.e. $N = 1$ (green), 
and continuing with periodic controls for $N = 2, 3, 4, \dots$ (black).
%Each curve is produced by varying the released infected pests $\widetilde P_I$ over ``all possible" values.
Each point on these curves corresponds to a positive value of the released infected pests $\widetilde P_I$.
Using Fig.~3a-c, we can find control strategies that give a profit close to optimal,
in the sense of one-off and periodic control strategies, and which are simultaneously efficacious in the sense of having a short half-biomass time.
This is possible since the slope of the Pareto front, in the sense of one-off and periodic control strategies,
(light green curve) is small near the maximum profit of the front, in particular for low to intermediate immigration rates.
We recommend such Pareto-efficient strategies in place of optimizing only profit, to obtain more stable outcomes.

In Table~\ref{tab:one}, we specify in detail four representative strategies on the Pareto front. These strategies are marked with circles in Fig.~3a-c. All four
strategies have a half-biomass time of less than 80 and profit not far from the highest that can be achieved for any strategy of the same type. Depending on how one wishes to trade off between profit and efficacy,
other strategies on the Pareto front with higher or lower efficacy can naturally also be considered.

\subsection{One-off control strategies are sufficient for low to intermediate immigration while periodic control is preferable for high immigration}
\label{sec:periodic}

To find the type of preferable control strategies for different immigration rates of susceptible pests, 
we assume, as a rule of thumb, that the farmer wants to keep the half-biomass time below 80. 
Further improvements in efficacy will be chosen only if the slope of the Pareto-front is very small near half-biomass time 80, 
so that efficacy can be substantially improved to the cost of hardly no profit.   

From Fig.~3a-c we can conclude that one-off control strategies are preferable for low to intermediate immigration rates, 
as such strategies can easily push the half-biomass time below 60 for low immigration and below 80 for intermediate immigration. 
For high immigration rates, we recommend periodic control with either two or three ($N = 2$ or $N = 3$) releases of infected pests.
This is because one-off control cannot achieve sufficiently small half-biomass time in this case, see Fig.~3c. 
In particular, in a field with high immigration,
using periodic control in place of one-off control can result in the half-biomass time being reduced by $25\%$ without reducing the profit.
This is a natural and expected result:
If the immigration of susceptible pests is high,
then it is not enough to ``push down" the pest biomass only in the beginning of the season,
a more continuous maintenance, such as periodic control, is needed.

%
% TABLE 2
%

\begin{table}[!hbp]
\caption{Representative pest-control strategies.}
\begin{center}
\begin{tabular}{@{}lllll}
  \hline
   Immigration, $A$          & Profit         & Half-biomass time  & Released infected   & Number of   \\
  (m$^{-2}$ day$^{-1}$) & ($\$$ m$^{-2}$)   & (days)            &  pests, $\widetilde P_I$ (m$^{-2}$) & pest bursts, $N$\\
  \hline
 $50$             & 0.132          &    53              & 64        & 1 (one-off)\\
  \hline
 $150$            & 0.126          &    72              & 256       & 1 (one-off)\\
%  \hline
% $200$            & 0.119          &    86              & 512       & 1\\
%  \hline
% $250$            & 0.087          &    106             & 2048 \\
%  \hline
  \hline
 $250$            & 0.089          &    79              & 1700        & 2 (periodic)\\
  \hline
 $250$            & 0.089          &    79              & 1450        & 3 (periodic)\\
  \hline
\label{tab:one}
\end{tabular}
\end{center}
\end{table}

%
% FIGURE 2
%

\begin{figure}
\begin{center}
  \includegraphics[scale=0.4]{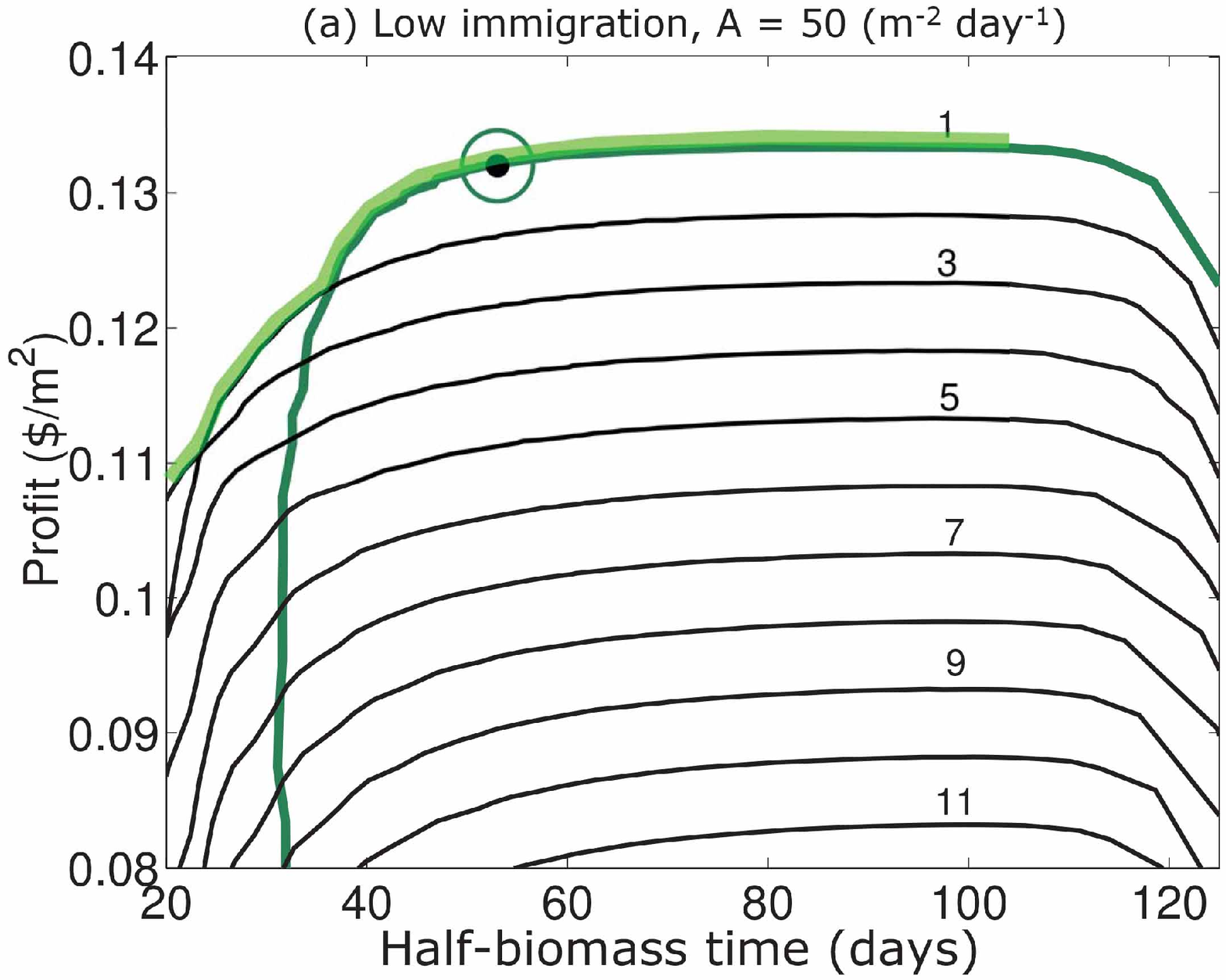}%\hspace{0.8cm}
  \includegraphics[scale=0.4]{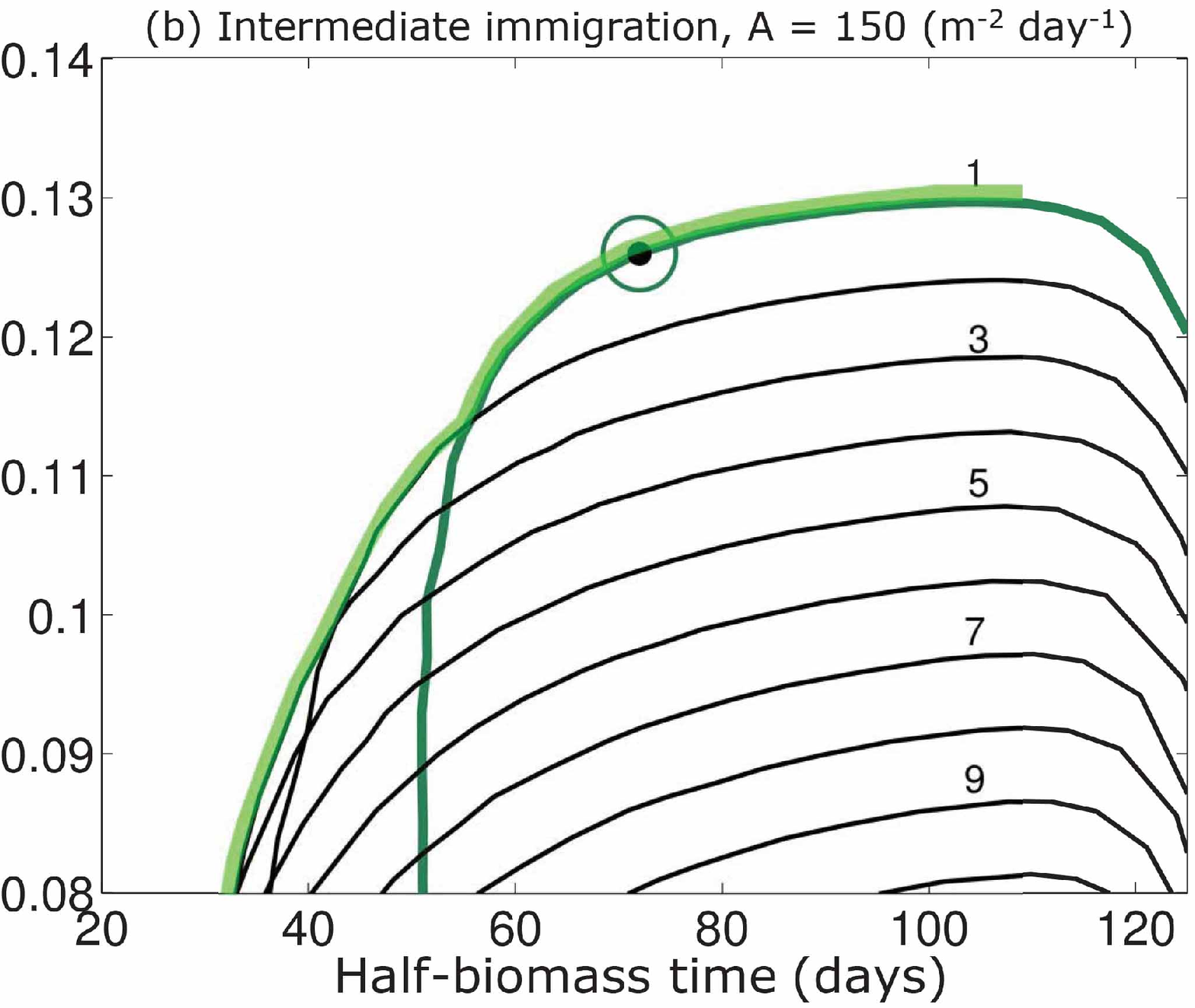}
  \includegraphics[scale=0.4]{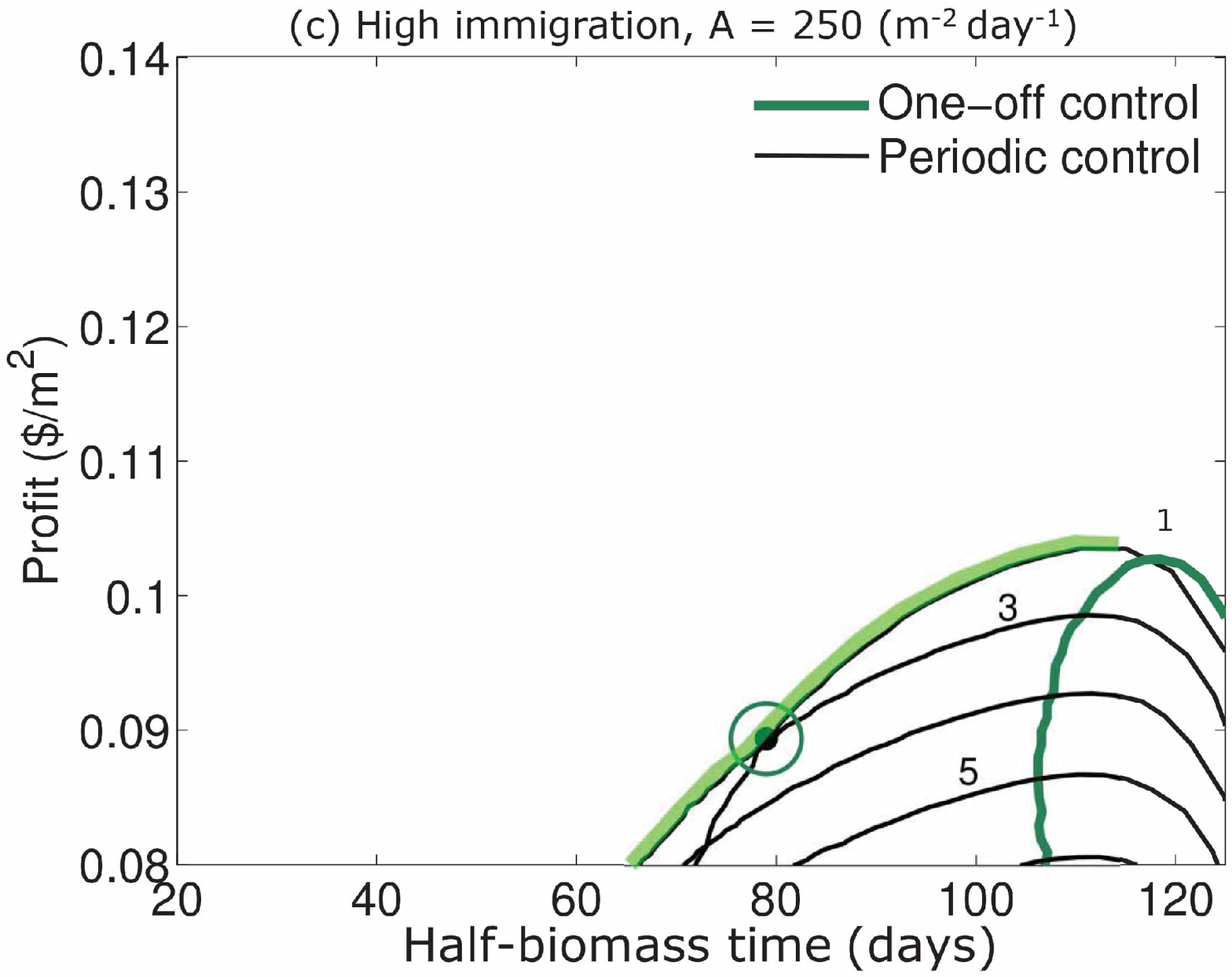}
  \includegraphics[scale=0.4]{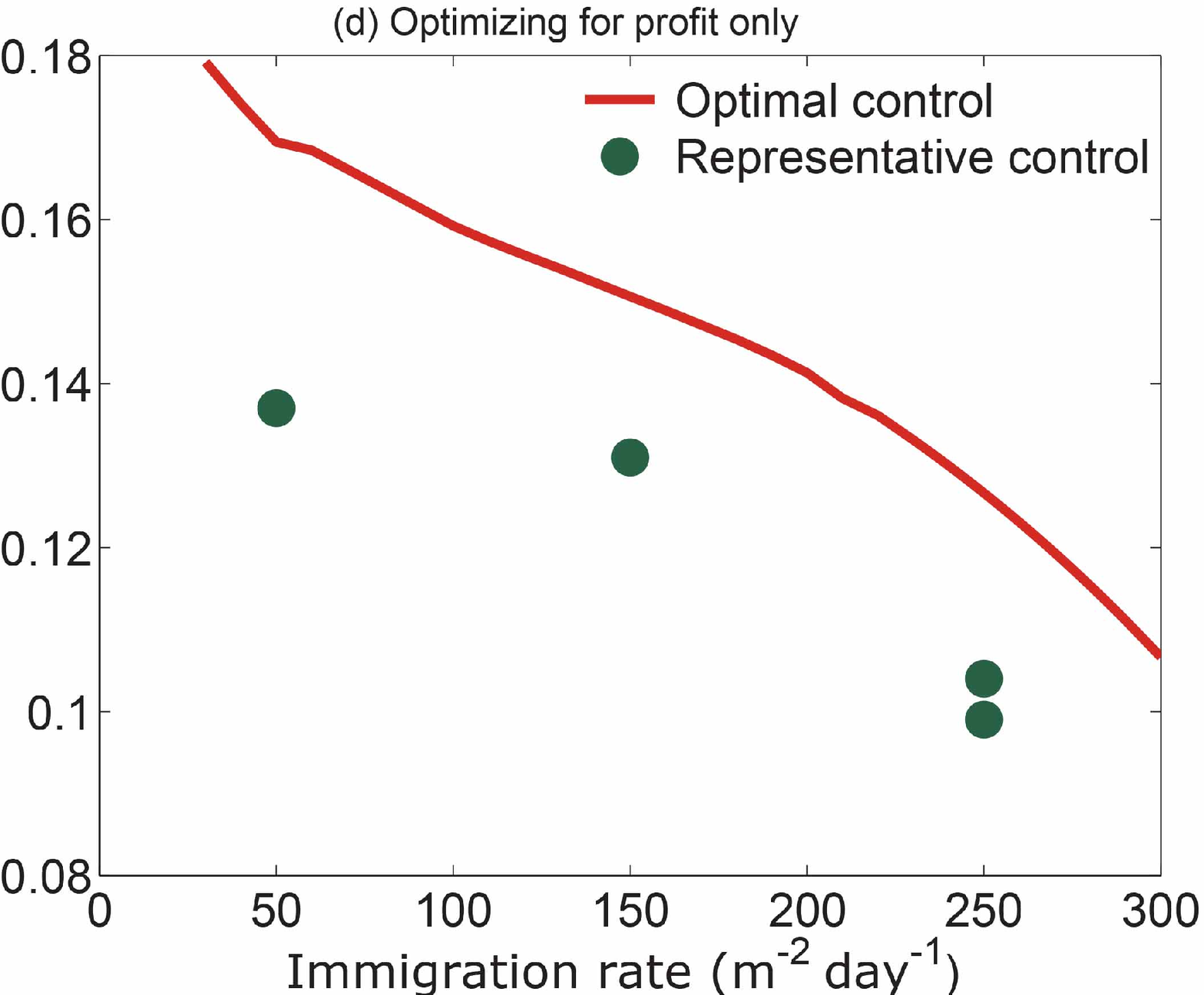}\hspace{0.4cm}
  \caption{Pareto-efficient control strategies allow safe controls to a small cost in profit. 
The circles indicate representative control strategies on the Pareto front (light green curve). 
In each subfigure, different curves correspond to different $N$, starting with one-off control (green), 
and continuing with periodic controls for $N = 2, 3, 4, \dots$ (black).
%Each curve is produced by varying the released infected pests $\widetilde P_I$ over ``all possible" values. 
Each point on these curves corresponds to a positive value of the released infected pests $\widetilde P_I$.
Panels (a) and (b): For low to intermediate immigration, we recommend one-off controls,
while for high immigration (c) periodic control is better and we recommend either two or three bursts ($N = 2$ or $N = 3$).
Note that at the center of the circle in panel (c) the periodic control with two bursts ($N = 2$)
coincides with the periodic control with three bursts ($N = 3$).
(d): The representative control strategies are not far from optimal in the sense of profit.
  The optimal profit obtained by running the software TOMLAB, maximizing only profit, is given by the red curve.
  The profit of the representative control strategies is marked with green balls.
  In panel (d), all results are without cost for labour, i.e. $p_{\text{labour}} = 0$.}
\end{center}
\label{fig:two}
\end{figure}

\subsection{The determined control strategies are not far from optimal in terms of profit}
\label{sect_optimal}

To compare the representative control strategies in Table~\ref{tab:one} to the optimal profit,
when considering profit as the only target under optimization, we determined the best possible
profit using the software TOMLAB \citep{KK99}. 
Unlike the control strategies we have considered, this optimal control strategy can take any form and may be very
hard to implement in practice. To compare the profit achieved through applying this optimal
control strategy with those of our representative control strategies, we assume in this section that
the labour cost for placing infected pests in the field is zero, i.e. $p_{\text{labour}} = 0$. This
assumption is necessitated by the fact that the determined optimal control strategy
may require continuous release of infected individuals. By making this assumption, we thus achieve
a situation in which the difference in profit between the two strategies should be largest.

Figure~3d shows the profit of the representative control strategies given in Table~\ref{tab:one}
(with $p_{\text{labour}} = 0$) together with the optimal profit obtained by TOMLAB.
We conclude that the profit of the representative control strategies are not far from the optimal profit in the whole range of immigration rate.
In particular, the representative control strategies yield more than $85\%$ of the optimal profit.
We remark that the gap between the optimal profit and the representative control strategies would decrease further when adding any reasonable labour cost to both controls.
We have also observed, not surprisingly, that the trajectories for crop resulting from the optimal control strategies obtained by simulations in TOMLAB have a large half-biomass time.
Therefore, they are sensitive to perturbations which in turn clearly yields an additional cost.

\subsection{Robustness}
\label{sect_robus}

Consequences of variations in the price parameters can be understood from
the profit function \eqref{eq:profit-func}.
In particular, increasing (decreasing) fixed costs $p_{\text{fixed}}$ will only lower (lift) all the curves in Fig.~3,
and therefore our results are robust with variations in $p_{\text{fixed}}$.
Increasing the price of placing infected pests in the field ($p_{\text{labour}}$) will be in favour of
controls with few pest bursts such as one-off control ($N = 1$).
To understand the magnitude of this dependence we note from the profit function \eqref{eq:profit-func} that
increasing (decreasing) $p_{\text{labour}}$
will lower (lift) the curves with $N \times$ the increase (decrease), where $N$ is the number of pest bursts.

We proceed by investigating robustness of our results by varying values of crop intrinsic growth rate $r$,
contact rate $\beta$ as well as immigration of susceptible pests $A$.
Figures~4a-c show regions where the recommended control is one-off control,
periodic control and where it is impossible to obtain a positive profit. %in the case of intermediate immigration rate, $A = 150$.
%Figure 4 (a) and (b) show similar simulations but for $r = 0.45$ and $\beta = 0.008$ fixed, respectively.
To find a border between recommending one-off or periodic control, we use here the following rule of thumb:
We recommend to use periodic control,
in place of one-off control,
if the half-biomass time thereby can be reduced by at least $20\%$ without loosing more than $20\%$ of the best profit resulting from the one-off controls.
Figures~4a-c are produced by examination of numerous
figures of the same type as Fig.~3a-c.
Besides results illustrated in Figs.~4a-c,
we concluded from these simulations that the shape of the curves are rather stable.
This means that our conclusion that a small reduction in profit allows efficacious control strategies is robust.
Moreover, our conclusion that one-off control strategies are sufficient for (relatively) low to intermediate immigration holds whenever $r > 0.4$ and $\beta > 0.007$.
In addition, the later conclusion can be extended as: one-off control strategies are sufficient for relatively low to intermediate immigration, as well as for relatively high contact rate and relatively high crop intrinsic growth rate.

The simulations in this section also verifies the natural facts that crop growth will increase in $r$, $\beta$ and decrease in $A$.

%\begin{figure}
%\begin{center}
%  \includegraphics[scale=0.4]{sensitivity_pest.eps}%\hspace{0.8cm}
%  \caption{Regions in the $\beta,r$-plane where the recommended control is one-off control (green),
%periodic control (grey) and where it is impossible to obtain a positive profit (red) in the case of intermediate immigration %rate, $A = 150$. The black dot represents parameter values $r = 0.45$ and $\beta = 0.008$ used in Figures 1 and 2.}
%\end{center}
%\label{fig:sens}
%\end{figure}

\begin{figure}
\begin{center} \hspace{0.1cm}
  \includegraphics[scale=0.4]{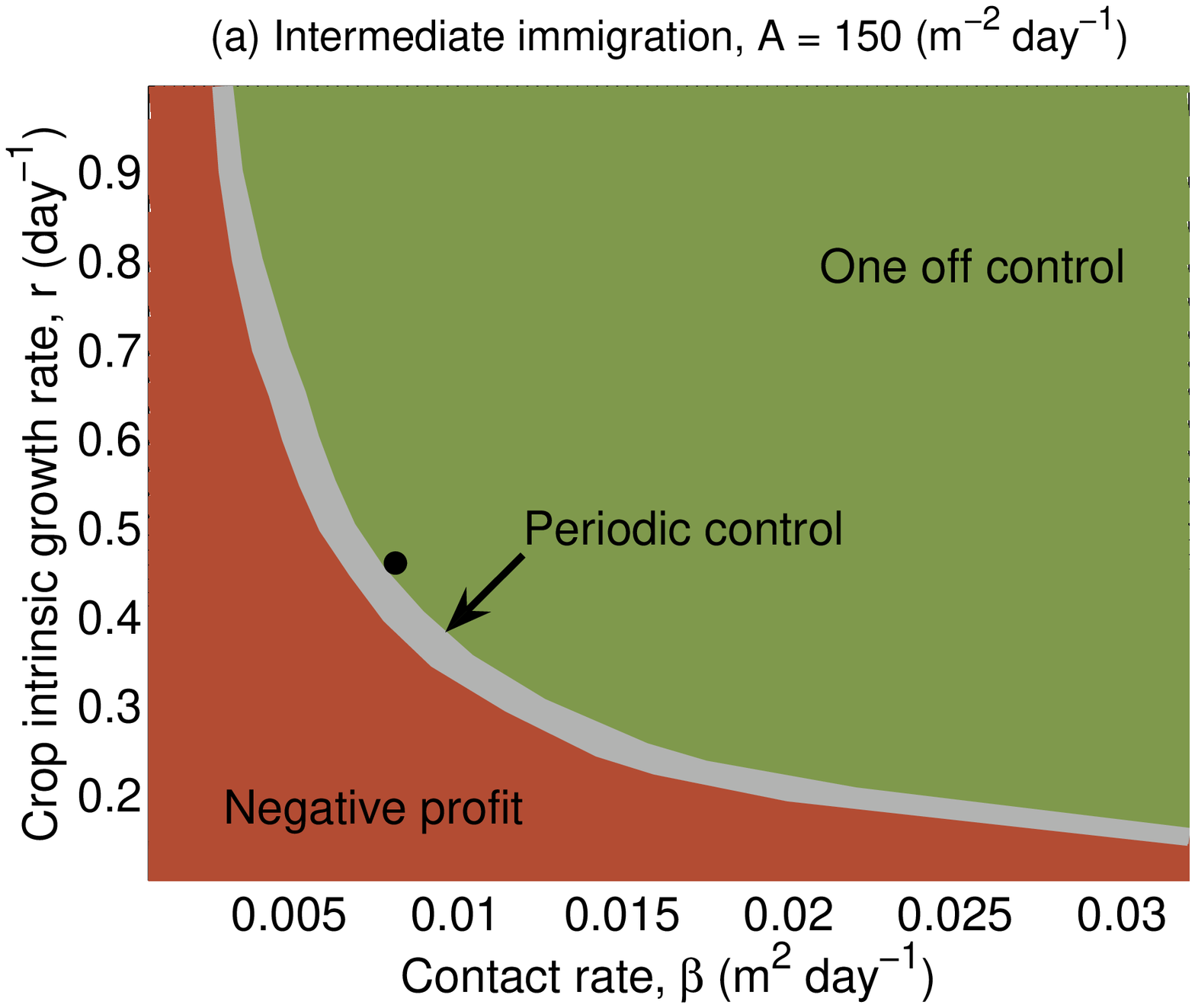}\hspace{0.1cm}
  \includegraphics[scale=0.4]{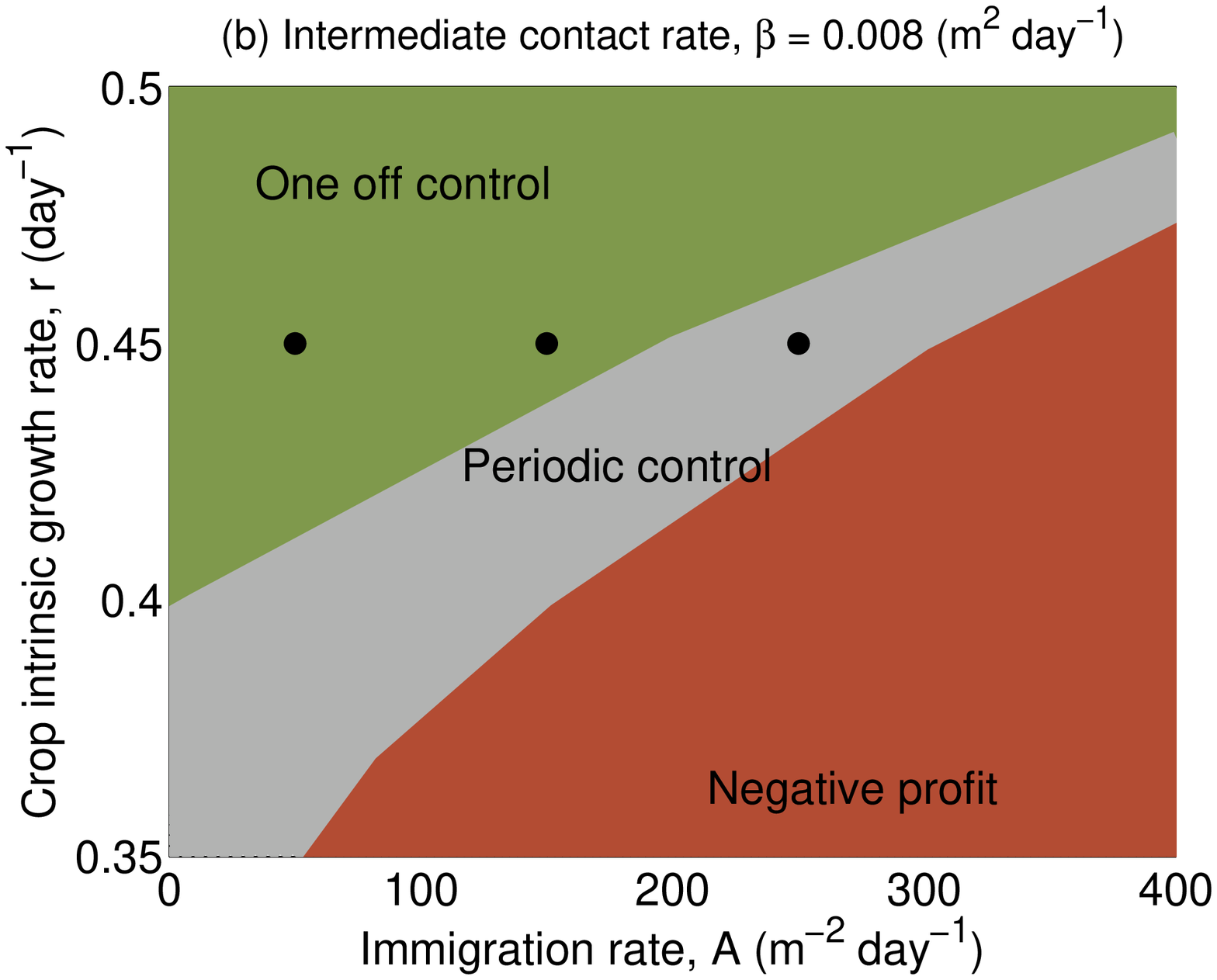}\hspace{-0.1cm}
  \includegraphics[scale=0.4]{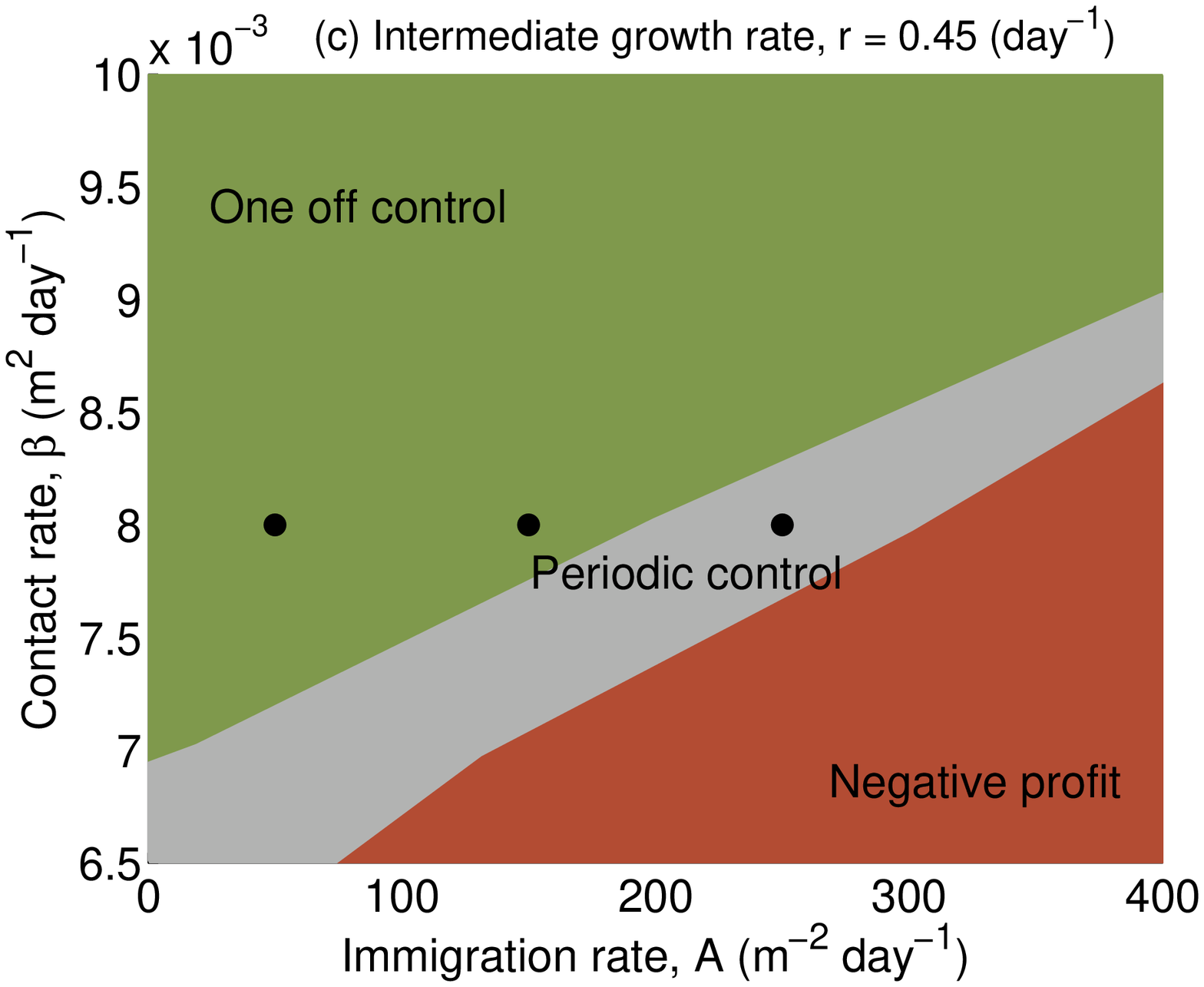}\hspace{-0.3cm}
  \includegraphics[scale=0.4]{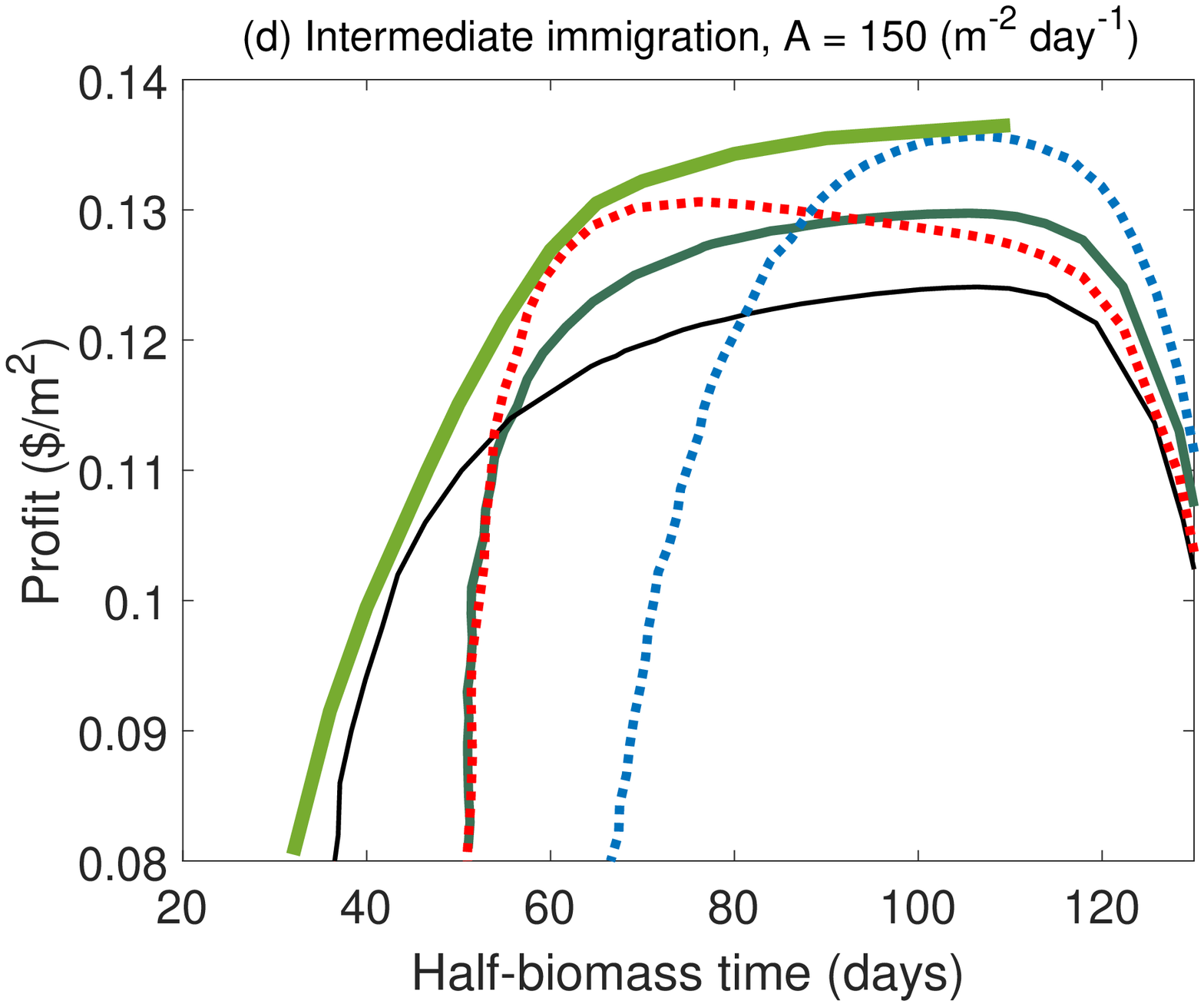}
  \caption{(a)-(c) Regions where the recommended control is one-off control (green),
periodic control (grey) and where it is impossible to obtain a positive profit (red).
The black dots represent parameter values used in Fig.~3.
In each subfigure, 100 different parameter combinations, equally distributed over the presented area, have been tested.
  (a) Immigration rate $A = 150$ is fixed while crop intrinsic growth rate $r$ and contact rate $\beta$ are varied.
  (b) Contact rate $\beta = 0.008$ is fixed while crop intrinsic growth rate $r$ and immigration rate $A$ are varied.
  (c) Crop intrinsic growth rate $r = 0.45$ is fixed while contact rate $\beta$ and immigration rate $A$ are varied.
  (d) Profit and half-biomass time when allowing for combinations of periodic controls and one-off controls in the setting of intermediate immigration $A = 150$.
  The light-green curve represents the Pareto front corresponding to ``all possible" combined control strategies.
  One-off control (dark-green),
  periodic control with $N = 2$ and $\tau = 7$ (black),
  periodic control with $N = 2$ and $\tau = 137$ (red, dotted), and
  combined one-off and periodic control consisting of
  one impulse $\widetilde{P}_{\text{one-off}}$ at time 0 and another impulse $\widetilde{P}_{\text{periodic}}$ at time
  137 with
  $\widetilde{P}_{\text{one-off}} = 0.1 \widetilde{P}_{\text{periodic}}$ (blue, dotted).
  }
\end{center}
\label{fig:sens_2}
\end{figure}

%%%%%%%%%%%%%%%%%%%%%%%%%%%%%%%%%%%%%%%%%%%%%%%%%%%%%%%%%%%%%%%%%%%%%%
%%%%%%%%%%%%%%%%%%%%%%%%%%%%%%%%%%%%%%%%%%%%%%%%%%%%%%%%%%%%%%%%%%%%%%
%%%%%%%%%%%%%%%%%%%%%%%%%%%%%%%%%%%%%%%%%%%%%%%%%%%%%%%%%%%%%%%%%%%%%%
%%%%%%%%%%%%%%%%%%%%%%%%%%%%%%%%%%%%%%%%%%%%%%%%%%%%%%%%%%%%%%%%%%%%%%
%%%%%%%%%%%%%%%%%%%%%%%%%%%%%%%%%%%%%%%%%%%%%%%%%%%%%%%%%%%%%%%%%%%%%%
%%%%%%%%%%%%%%%%%%%%%%%%%%%%%%%%%%%%%%%%%%%%%%%%%%%%%%%%%%%%%%%%%%%%%%
%%%%%%%%%%%%%%%%%%%%%%%%%%%%%%%%%%%%%%%%%%%%%%%%%%%%%%%%%%%%%%%%%%%%%%
%%%%%%%%%%%%%%%%%%%%%%%%%%%%%%%%%%%%%%%%%%%%%%%%%%%%%%%%%%%%%%%%%%%%%%
%%%%%%%%%%%%%%%%%%%%%%%%%%%%%%%%%%%%%%%%%%%%%%%%%%%%%%%%%%%%%%%%%%%%%%
%%%%%%%%%%%%%%%%%%%%%%%%%%%%%%%%%%%%%%%%%%%%%%%%%%%%%%%%%%%%%%%%%%%%%%

%\newpage

\section{Discussion}

We have considered biological control of agricultural pests. Using a
dual-objective approach and the economic concept of Pareto efficiency,
we have determined one-off and periodic control strategies that are
stable to perturbations and simultaneously nearly optimal in terms of
profit.
Our optimization approach as well as our measure of efficacy are general
and can be applied to effectively any crop-pest-pathogen system. %s beyond system \eqref{eq:syst_eq_1}-\eqref{eq:syst_eq_3} as well.
Depending on the immigration rate of pests from nearby fields,
we recommend either one-off control, with entomopathogens released
only once in the beginning of the growing season, or periodic control,
with entomopathogens released at periodic intervals. Surprisingly, as
we showed in Section~\ref{sect_optimal}, these two conceptually simple
pest-control strategies come close to the best that can be
achieved, even when allowing for complicated continuous-release
strategies.

The threshold for the immigration rate at which one should
switch from one-off control to periodic control as well as the other
control parameters can be determined through our approach after a model has been selected and 
parameterized for the system of interest.
We also believe that it should be possible to determine reasonable values for the control
parameters through controlled field experiments, or even individual
experimentation. Future work may aim to overcome the problem of
parametrization by deriving even more general insights, such as
adaptive rules that relate the timing of release or the released
quantity to observed changes in the crop and pest population.

When defining the periodic control strategies in
Section~\ref{sec:model_control}, we assumed that the same amount of pests
is released once a week. This assumption should be considered more as
an example than a rule. In fact, the efficacy of the periodic control
will only be marginally affected by smaller changes in the length of
this time interval. The costs associated with the periodic control
does, however, strongly depend on the price of placing the infected
pests in the field ($p_{\text{labour}}$). A lower price will make the
controls that releases more often (higher $N$) better. Moreover, a
natural extension of the periodic controls tested here would be to
allow for the combination of one-off control and periodic control. We
have seen that one-off control performs well in most tested cases, and
therefore we believe in releasing larger amounts of pests at impulses
in the beginning of the growth season, and smaller amounts later
during the season. We will further discuss this below.

We have considered dual-objective optimization which accounts for both
efficacy and profit. The fundamental drawback with optimizing profit
alone is that the resulting crop trajectory may behave similar to
the trajectory for crop biomass given in Fig.~\ref{fig:one}b. Even
worse, the increase in crop abundance may come closer to the end of
the season so that a small perturbation during the growth season may
pass it beyond the end of the season, as the trajectories of crop in
Fig.~\ref{fig:one}c and d.  Hence, instead of optimizing only
profit, farmers should consider a trade-off between profit and some
measure of efficacy (sensitivity to perturbations). To exclude growth patterns
that are very sensitive to perturbations and hence implies a high
risk for farmers, we introduced in Section~\ref{sec:dual} a measure
of efficacy, the half-biomass time, and applied a dual-objective
approach through Pareto efficiency.

Our measure of efficacy, half-biomass time, is related to resilience,
which is nowadays frequently used in ecology \citep{PL77, LB99, PCJMP02, MPS06, L10, VR-JG-NUD10}. 
Resilience can be measured as the reciprocal of the returntime of a trajectory to a stable
equilibrium, perturbed from the same equilibrium.
See Lundstr\"om et al. (2016, 2017) for definitions, 
discussions and applications of such nonlocal resilience measures.
In essence, short half-biomass time corresponds to high resilience as it means that the crop is quickly
able to approach an equilibrium state. 
We define half-biomass time based on typical growth patterns for crops in our model. 
However, a similar definition will apply to other models
showing similar growth patterns, involving models where biomass is
assumed to grow logistically. 
To stress the generality of our work, 
we note that the measure
half-biomass time may be replaced by any other suitable measure of efficacy which
can be defined for the model in question, 
and a similar analysis using Pareto-efficiency and simple control strategies can be carried out.

The study closest to ours is the effort by %\citet{card}
Cardoso et al. (2009) in
understanding biological control of caterpillars (\emph{Anticarsia gematalis})
by natural enemies such as wasps and spiders. Similar to us, these
authors use a dual-objective approach and use the economic concept of
Pareto efficiency. In their case, the two objectives are measures of
the number of prey (the pest) and the number of predators (the natural
enemy of the pest) which are released. Like us, these authors aim to
derive more practical pest-control strategies by moving away from
continuous-release strategies towards release at specified time
points, so-called impulsive release \citep{tang2005, zhang2007}. Apart
from the different choice of study system, a few additional differences are worth
noting: First, %\citet{card}
Cardoso et al. (2009) do not compare %attempt to further simplify '
the impulsive-control strategies that they derive
%.
%Specifically, they do
%not
%compare the difference between the optimal strategies and
with simple one-off control and periodic-control strategies as considered
here. Second, whereas we develop and motivate a non-linear measure of
pest-control efficacy, %\citet{card}
Cardoso et al. (2009) uses a measure of the number of
released natural enemies as their second objective.
Third, we explicitly
consider the dynamics of the crop and move beyond the traditional
Lotka-Volterra framework by incorporating functional
responses. Interestingly, the optimal strategies presented in
Figs.~3-5 of %\citet{card}
Cardoso et al. (2009) have a clear resemblance to a periodic
control except at the beginning and the end of season. Based on this
observation and our own results, we conjecture that effective
strategies of biological pest control
%even more effective than the representative controls in Table 2,
can be obtained by combining one-off control and periodic control.
Figure~4d shows the Pareto-front for profit and half-biomass time when
allowing for such combinations in the setting of intermediate immigration $A = 150$.
In particular, the combined control has four control variables;
the amount of released pests in the beginning of the season ($\widetilde{P}_{\text{one-off}}$),
the total amount of released pests during the following periodic control ($\widetilde{P}_{\text{periodic}}$),
the number of days between pest impulses ($\tau$) and
the total number of impulses (${N}$).
By comparing Fig.~3b with Fig.~4d we conclude that the Pareto front is slightly
higher in Fig.~4d.
Thus, allowing for combinations of periodic controls and one-off controls opens for
slightly more efficient control strategies.
This is expected since such approach allows for more degrees of freedom and can therefore better meet the
populations need of pest control.
Note, however, that we do not obtain a substantially better result.
Moreover, the dotted curves in Fig.~4d,
which reaches near Pareto optimality,
consists of strategies with only two impulses ($N = 2$),
the first at the beginning of the season and the second near the end of the season.
Hence, simple strategies performs well also when allowing for combinations of one-off control and periodic control.
%(a) initial
%amount to be released and delay until onset of periodic control, (b)
%periodic amount to be released and interval between two successive
%releases, and (c) the time before the end of the season when
%pest-control efforts ceases.

As noted in this paper, several studies in applied mathematics have
aimed to determine optimal strategies of biological pest
control. These studies draw upon and synthesize a rich scientific
heritage of mathematical modeling in ecology, dynamical systems
theory, and optimal control theory. The crowning achievement is the
ability to determine the optimal timing for releasing natural enemies
across a range of pest-enemy systems. 
This accomplishment will, however, be of limited value if insights are not
distilled and results disseminated in a form that is useful for
the agricultural community.
Our aim with this paper has been to
demonstrate how practically useful pest-control strategies can be
determined from mathematical models of crop-pest-enemy
interactions.
While we have arguably not succeeded in bridging the
full width of the gulf that currently exist between mathematical theory
and practical applications, we expect this gulf to shrink
as future authors continue our effort to uncover practically useful
strategies of biological pest control. \\

%%%%%%%%%%%%%%%%%%%%%%%%%%%%%%%%%%%%%%%%%%%%%%%%%%%%%%%%%%%%%%%%%%%%%%
%%%%%%%%%%%%%%%%%%%%%%%%%%%%%%%%%%%%%%%%%%%%%%%%%%%%%%%%%%%%%%%%%%%%%%
%%%%%%%%%%%%%%%%%%%%%%%%%%%%%%%%%%%%%%%%%%%%%%%%%%%%%%%%%%%%%%%%%%%%%%
%%%%%%%%%%%%%%%%%%%%%%%%%%%%%%%%%%%%%%%%%%%%%%%%%%%%%%%%%%%%%%%%%%%%%%
%%%%%%%%%%%%%%%%%%%%%%%%%%%%%%%%%%%%%%%%%%%%%%%%%%%%%%%%%%%%%%%%%%%%%%
%%%%%%%%%%%%%%%%%%%%%%%%%%%%%%%%%%%%%%%%%%%%%%%%%%%%%%%%%%%%%%%%%%%%%%
%%%%%%%%%%%%%%%%%%%%%%%%%%%%%%%%%%%%%%%%%%%%%%%%%%%%%%%%%%%%%%%%%%%%%%
%%%%%%%%%%%%%%%%%%%%%%%%%%%%%%%%%%%%%%%%%%%%%%%%%%%%%%%%%%%%%%%%%%%%%%
%%%%%%%%%%%%%%%%%%%%%%%%%%%%%%%%%%%%%%%%%%%%%%%%%%%%%%%%%%%%%%%%%%%%%%
%%%%%%%%%%%%%%%%%%%%%%%%%%%%%%%%%%%%%%%%%%%%%%%%%%%%%%%%%%%%%%%%%%%%%%

\noindent
{\bf Acknowledgement}.
The work of HZ was supported by the Swedish Research Council,
the National Natural Science Foundation of China,
Grant ID 11201187 and the Scientific Research Foundation for the Returned Overseas Chinese Scholars and the China Scholarship Council.
The authors thank Daniel Simpson for useful discussions and comments.\\

\end{document}